\documentclass[aps,prb,twocolumn,showpacs,10pt]{revtex4-1}
\pdfoutput=1
\usepackage{graphicx} 
\begin{document}

\newcommand{\AFM}{\text{AFM}}
\newcommand{\eff}{\text{eff}}
\newcommand{\eps}{\varepsilon}
\newcommand{\Dv}{\mathbf D}
\newcommand{\Sv}{\mathbf S}
\newcommand{\sumij}{\langle ij\rangle}

\title{Spin-chain magnetism and uniform Dzyaloshinsky-Moriya anisotropy in BaV$_3$O$_8$}

\author{Alexander A. Tsirlin}
\email{altsirlin@gmail.com}
\affiliation{National Institute of Chemical Physics and Biophysics, 12618 Tallinn, Estonia}


\begin{abstract}
We report on the microscopic magnetic model of a spin-$\frac12$ magnet BaV$_3$O$_8$. In contrast to earlier phenomenological analysis, our density-functional band-structure calculations combined with quantum Monte-Carlo simulations establish a relatively simple and non-frustrated model of weakly coupled spin chains with intrachain coupling $J\simeq 38$~K and N\'eel temperature $T_N\simeq 6$~K, both in excellent agreement with experiment. The intrachain coupling between spin-$\frac12$ V$^{+4}$ ions takes place via two contiguous V$^{+5}$O$_4$ tetrahedra forming an extended superexchange pathway with the V$^{+4}$--V$^{+4}$ distance of 7.44~\r A. Surprisingly, this pathway is preferred over shorter V$^{+4}$--V$^{+4}$ connections, owing to peculiarities in the interacting orbitals of the magnetic V$^{+4}$ ions and V$^{+5}$ ions that are non-magnetic, but feature low-lying $3d$ states contributing to the superexchange process. We also note that the crystal structure of BaV$_3$O$_8$ supports the long-sought uniform arrangement of Dzyaloshinsky-Moriya (DM) couplings on a spin-$\frac12$ chain. While our calculations yield only a weak DM anisotropy in BaV$_3$O$_8$, the crystal structure of this compound provides a suitable framework for the search of spin chains with the uniform DM anisotropy in other compounds of the vanadate family.     
\end{abstract}
\maketitle

\section{Introduction}
Quantum spin systems reveal exotic ground states and facilitate experimental access to many-body quantum phenomena.\cite{balents2010,han2012} These interesting effects are enhanced in systems with strong quantum fluctuations, which are typically driven by the magnetic frustration, i.e., the competition between different magnetic couplings. The identification of frustrated spin systems remains an important, but also non-trivial problem in solid-state physics. The conventional criterion\cite{greedan2001} of the Curie-Weiss temperature to the N\'eel temperature ($\theta/T_N$) ratio being high (typically, $\theta/T_N\geq 10$) is somewhat ambiguous, because the relation between the magnetic ordering temperature $T_N$ and the overall energy scale of the magnetic couplings $\theta$ reflects only the tendency of the system to avoid long-range magnetic order. This tendency may be driven by several effects, such as frustration and low-dimensionality. When the spin lattice comprises weakly coupled spin chains or spin planes, the magnetic ordering temperature may be low as well, even without the frustration.\cite{yasuda2005}

\begin{figure*}
\includegraphics{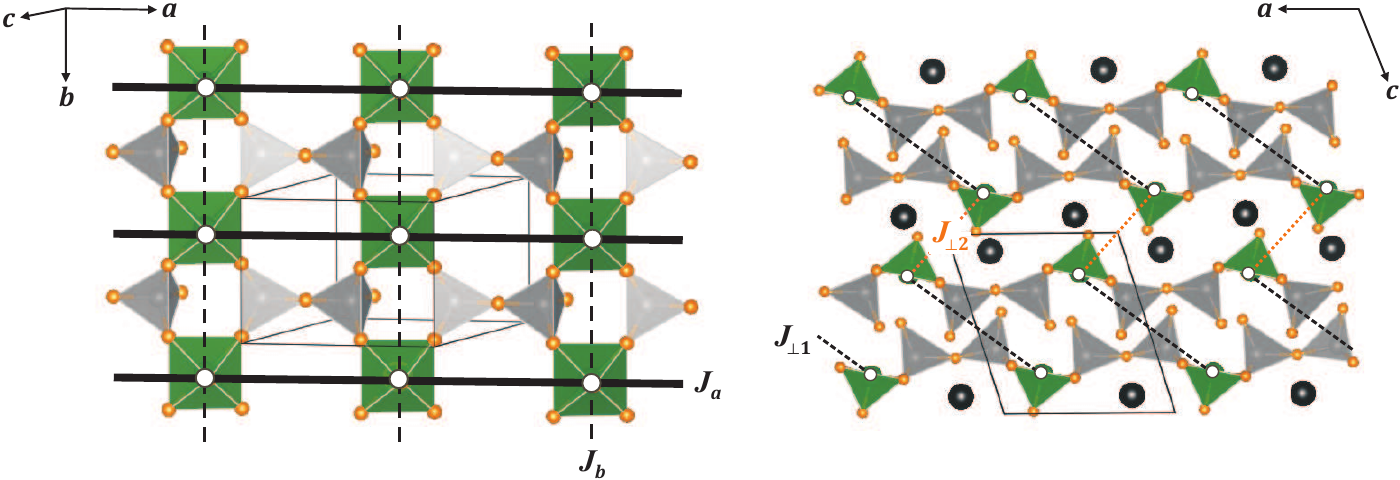}
\caption{\label{fig:str}
(Color online) Crystal structure of BaV$_3$O$_8$. Left panel: the [VOV$_2$O$_7$] layer comprising V$^{+4}$O$_5$ pyramids (green, V2) and V$^{+5}$O$_4$ tetrahedra (gray, V1 and V3). Right panel: the stacking of layers into bilayers separated by the Ba cations (black spheres). Leading intralayer interactions $J_a$ (spin chains, solid line) and $J_b$ (dashed lines) as well as the interlayer couplings $J_{\perp1}$ (within the bilayer, short-dashed lines) and $J_{\perp2}$ (between the bilayers, dotted orange lines) are shown. Crystal structures are visualized using the \texttt{VESTA} software.\cite{vesta}
}
\end{figure*}
Here, we focus on the spin-$\frac12$ compound BaV$_3$O$_8$ that was recently identified as a frustrated spin-chain magnet, following its $\theta/T_N$ ratio of 5. According to Chakrabarty \textit{et al.},\cite{chakrabarty} the spin model of BaV$_3$O$_8$ can be described as a Majumdar-Ghosh chain with a $J_2/J_1$ ratio of 2. This model is better known as the $J_1-J_2$ spin chain and entails competing nearest-neighbor and next-nearest-neighbor couplings $J_1$ and $J_2$, respectively. We will use density-functional (DFT) band-structure calculations and ensuing numerical simulations to propose an alternative magnetic model of non-frustrated spin chains with the intrachain coupling of $J\simeq 38$~K in excellent agreement with the experiment. Our model is also able to reproduce the experimental magnetic ordering temperature $T_N\simeq 6$~K without invoking any magnetic frustration. Moreover, our model emphasizes another -- and so far unknown -- feature of BaV$_3$O$_8$, the formation of spin chains with the uniform Dzyaloshinsky-Moriya (DM) anisotropy, which has been long sought in spin-$\frac12$ magnetic materials.\cite{starykh2008,garate2010}

BaV$_3$O$_8$ is a charge-ordered vanadium-based compound comprising V$^{+4}$ and V$^{+5}$ ions that show drastically different local environment (Fig.~\ref{fig:str}).\cite{oka1995,marsh1996} The \mbox{spin-$\frac12$} V$^{+4}$ ion occupies the crystallographic position V2, which is octahedrally coordinated. The non-magnetic V$^{+5}$ ions are in the tetrahedrally coordinated positions V1 and V3. The polyhedron of V$^{+4}$ shows a strong axial distortion: the V2--O6 bond is shortened to 1.63~\r A (vanadyl bond), the V2--O2 bond extends up to 2.18~\r A, while the four remaining bonds in the equatorial plane are at $1.95-2.0$~\r A. This peculiarity of the local environment has an immediate effect on the crystal-field levels (see also Fig.~\ref{fig:dos}) and imposes half-filling of the $d_{xy}$ orbital, where $xy$ is the equatorial plane of the octahedron.\cite{kaul2003,tsirlin2008} The O2 and O6 atoms do not lie in this $xy$ plane and do not take part in the magnetic superexchange.\cite{kaul2003,tsirlin2008} Therefore, it is more convenient to consider the polyhedron of V2 as a square pyramid, with the O6 atom removed.

The crystal structure of BaV$_3$O$_8$ visualized in terms of the V$^{+4}$O$_5$ pyramids and V$^{+5}$O$_4$ tetrahedra is shown in Fig.~\ref{fig:str}. It entails [VOV$_2$O$_7$] layers built by the pyramids and tetrahedra. The connections between the neighboring pyramids along $b$ are provided by double bridges of the V$^{+5}$O$_4$ tetrahedra, whereas longer connections along $a$ go through two contiguous tetrahedra (Fig.~\ref{fig:str}, left). Two neighboring layers form a bilayer. The bilayers separated by the Ba$^{+2}$ cations are stacked on top of each other along $c$.

\section{Method}  
In order to understand the magnetism of BaV$_3$O$_8$ on the microscopic level, we perform DFT band-structure calculations and evaluate individual exchange couplings. The calculations are done in the framework of the local density approximation (LDA)\cite{pw92} and the generalized gradient approximation (GGA)\cite{pbe96} in the full-potential local-orbital code \texttt{FPLO}\cite{fplo} and projector-augmented-wave code \texttt{VASP5.3}.\cite{vasp1,*vasp2} Reciprocal space was sampled with 518~$k$-points for the crystallographic unit cell (symmetry-irreducible part of the first Brillouin zone), 64~$k$-points for the two-fold supercells (48~atoms) used in the evaluation of isotropic exchanges, and 18~$k$-points for the four-fold supercell (96~atoms) used in the evaluation of anisotropy parameters. The convergence with respect to the $k$-mesh was carefully checked.  

Vanadium atoms entail strongly correlated $3d$ electrons that are not properly treated within LDA or GGA. We circumvented this problem by two different methods. First, we mapped relevant LDA bands onto a tight-binding model and further onto a Hubbard model that in the case of half-filling and strong correlations can be reduced to the Heisenberg model with the antiferromagnetic (AFM) exchange couplings $J_i^{\AFM}=4t_i^2/U_{\eff}$, where $t_i$ are hopping parameters derived from the LDA bands and $U_{\eff}=4$~eV is the effective on-site Coulomb repulsion on the vanadium site.\cite{tsirlin2011b,tsirlin2011c} Alternatively, we obtained total exchange couplings $J_i$ by mapping energies of collinear spin configurations onto the classical Heisenberg model. The total energies were obtained within GGA+$U$\footnote{We use the fully-localized-limit double-counting correction, which is common to \texttt{FPLO} and \texttt{VASP}.} with the on-site Coulomb repulsion $U_d=3$~eV (\texttt{FPLO}), $U_d=4$~eV (\texttt{VASP}) and on-site Hund's exchange $J_d=1$~eV (both codes).\cite{korotin1999,*korotin2000,tsirlin2011,tsirlin2011b,tsirlin2009} Here, we had to adjust the value of $U_d$ according to different basis sets used in the two codes.

We have also evaluated magnetic anisotropy parameters of the general spin Hamiltonian:
\begin{equation}
 \hat H=\sum_{\sumij}J_{ij}\Sv_i\Sv_j+\sum_{\sumij}\Dv_{ij}[\Sv_i\times\Sv_j],
\label{eq:ham}\end{equation}
where $\sumij$ denotes the summation over all bonds of the spin lattice, and $\Dv_{ij}$ are DM vectors (antisymmetric anisotropic exchange). The procedure is similar to that of the GGA+$U$ evaluation of $J_i$, but we have to use four-fold supercells, perform calculations with the spin-orbit coupling, and consider non-collinear spin configurations as well.\cite{xiang2011,anisotropy,supplement}

Thermodynamic properties of the microscopic magnetic model were obtained by quantum Monte-Carlo (QMC) simulations using the \texttt{loop} algorithm\cite{loop} of the \texttt{ALPS-1.3} simulation package.\cite{alps} We performed simulations for $L\times L/2\times L/4$ finite lattices with $L\leq 32$ and periodic boundary conditions. The lattice shape was adjusted\cite{sandvik1999} in order to match different correlation lengths of a quasi-one-dimensional (1D) spin system of BaV$_3$O$_8$.

\section{Results}
We start by comparing different structural data reported in the literature. The crystal structure of BaV$_3$O$_8$ was initially refined in the monoclinic space group $P2_1$ (Ref.~\onlinecite{oka1995}) and later re-determined in the higher-symmetry space group $P2_1/m$ (Ref.~\onlinecite{marsh1996}). In Ref.~\onlinecite{chakrabarty}, yet another refinement in the $P2_1/m$ group has been reported. We find the lowest energy for the original $P2_1/m$ structure. The $P2_1$ structure lies 0.16~eV/f.u. higher in energy. Therefore, the centrosymmetric space group $P2_1/m$ should be indeed preferred over its non-centrosymmetric subroup $P2_1$. The data from Ref.~\onlinecite{chakrabarty} appear to be very inaccurate and lead to the 8.49~eV/f.u. higher total energy. Indeed, these data reveal a skewed three-fold coordination of V1 and very high refinement residuals,\cite{chakrabarty} which are well above the standard reliability threshold.

\begin{figure}
\includegraphics{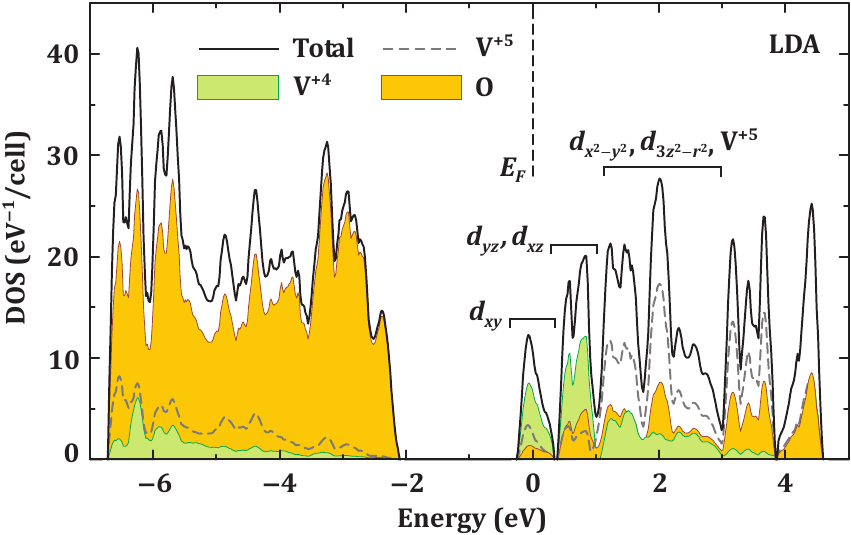}
\caption{\label{fig:dos}
(Color online) LDA density of states for BaV$_3$O$_8$. The Fermi level ($E_F$) is at zero energy. Crystal-field levels of V$^{+4}$ (crystallographic position V2) are labeled. The V$^{+5}$ states (crystallographic positions V1 and V3) dominate above 1~eV, only.
}
\end{figure}

In the following, we will only consider the $P2_1/m$ structure, as reported in Ref.~\onlinecite{marsh1996}. The relevant LDA band structure is shown in Fig.~\ref{fig:dos} and features O $2p$ bands between $-7$~eV and $-2$~eV followed by the V~$3d$ bands above $-0.5$~eV. We can clearly see the difference between the octahedrally-coordinated V$^{+4}$ (V2) and tetrahedrally-coordinated V$^{+5}$ (V1 and V3). The V2 states dominate the bands crossing the Fermi level. By contrast, the states of V$^{+5}$ are mostly found above 1~eV. Note that the LDA band structure is metallic, because the LDA fails to account properly for strong correlation effects in the V $3d$ shell. By using the LSDA+$U$ or GGA+$U$ methods, we are able to restore the insulating state with a typical band gap of 2.0~eV at $U_d=3-4$~eV.

\begin{figure}
\includegraphics{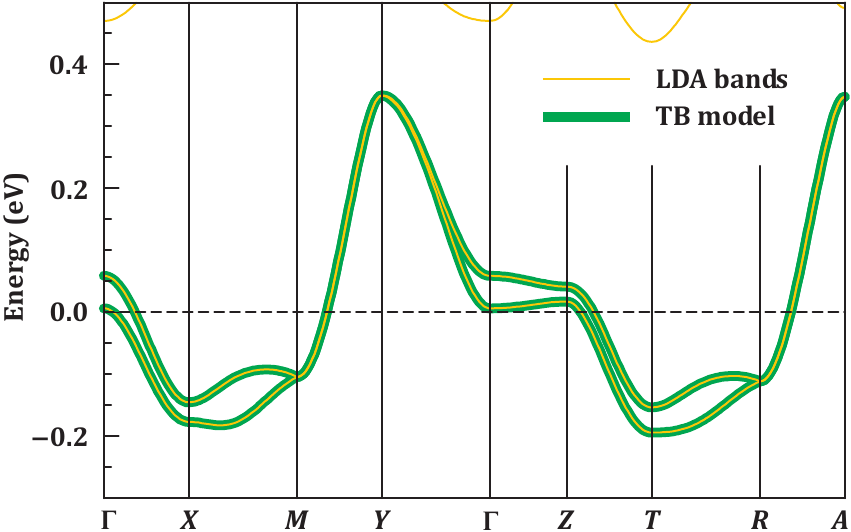}
\caption{\label{fig:bands}
(Color online) LDA band structure of BaV$_3$O$_8$ (thin light lines) and the fit with the tight-binding (TB) model for the $d_{xy}$ states of V$^{+4}$ (thick dark lines). The $k$ path is defined as follows: $\Gamma(0,0,0)$, $X(\frac12,0,0)$, $M(\frac12,\frac12,0)$, $Y(0,\frac12,0)$, $Z(0,0,\frac12)$, $T(\frac12,0,\frac12)$, $R(\frac12,\frac12,\frac12)$, $A(0,\frac12,\frac12)$.
}
\end{figure}

\begin{table}
\caption{\label{tab:exchange}
Magnetic couplings in BaV$_3$O$_8$: V$^{+4}$--V$^{+4}$ distances $d_i$ (in~\r A), hoppings $t_i$ (in~eV), AFM parts of the exchange integrals $J_i^{\AFM}=4t_i^2/U_{\eff}$ (in~K), and total exchange integrals $J_i$ (in~K) obtained from GGA+$U$ calculations with $U_d=4$~eV, $J_d=1$~eV (\texttt{VASP} code).
}
\begin{ruledtabular}
\begin{tabular}{ccrrc}
             & $d_i$(V$^{+4}$--V$^{+4}$) & \multicolumn{1}{c}{$t_i$} & $J_i^{\AFM}$ & $J_i$ \\
 $J_a$       &    7.43           & 0.082    &     78       &   38  \\     
 $J_b$       &    5.55           & $-0.042$ &     21       &    6  \\
 $J_{ab}$    &    9.28           & $-0.015$ &      3       &   --  \\ 
 $J_{\perp1}$&    7.78           & $-0.011$ &      1       &   --  \\
 $J_{\perp2}$&    5.50           & $-0.003$ &    0.1       &   --  \\
\end{tabular}
\end{ruledtabular}
\end{table}
A closer look at the V$^{+4}$ bands reveals crystal-field levels, which are expected for a $3d$ ion in a distorted octahedral environment. The order of levels is as follows: $\eps_{xy}<\eps_{yz}\simeq\eps_{xz}<\eps_{x^2-y^2}<\eps_{3z^2-r^2}$, where we use the local coordinate frame with the $z$ axis directed along the short V--O bond in the axial position of the octahedron, while $x$ and $y$ follow equatorial bonds in the basal plane, but are kept perpendicular to $z$. A decent description of the magnetism can be obtained from an effective one-orbital model including only the half-filled $d_{xy}$ states (Fig.~\ref{fig:bands}). An extended model with three orbitals can be constructed as well,\cite{mazurenko2006,tsirlin2011,tsirlin2011b,tsirlin2011c} but in our case it does not provide any additional information, because the hoppings from the half-filled $d_{xy}$ orbital to the empty $d_{yz}$ and $d_{xz}$ orbitals are negligibly small. The $d_{3z^2-r^2}$ and $d_{x^2-y^2}$ orbitals lie above 1.0~eV and strongly hybridize with the V$^{+5}$ states (see Fig.~\ref{fig:dos}) that hinder the reduction to an effective model containing V$^{+4}$ orbitals, only.  

Relevant hoppings between the $d_{xy}$ states are listed in Table~\ref{tab:exchange}. The unit cell of BaV$_3$O$_8$ features two V$^{+4}$ ions from two different layers (Fig.~\ref{fig:str}, right). Owing to the weak interlayer coupling, the $d_{xy}$ bands are close to double degeneracy (Fig.~\ref{fig:bands}). They show a strong dispersion along the $\Gamma-Y$ and $Y-M$ directions, which could imply sizable hoppings along both $a$ and $b$. However, a quantitative analysis using Wannier functions\cite{wannier} suggests that the leading AFM coupling is $J_a$ running along the $a$ direction, while the coupling $J_b$ along $b$ is much weaker (Table~\ref{tab:exchange}). The interlayer couplings follow the structural bilayers. The coupling within the bilayer $J_{\perp1}$ is stronger than the coupling between the bilayers $J_{\perp2}$. A weak frustration may be induced by the diagonal coupling $J_{ab}$ in the $ab$ plane, but its low value ($J_{ab}^{\AFM}\ll J_b^{\AFM}$) indicates a minor role of this frustration in the microscopic magnetic model. Indeed, even a non-frustrated spin model provides quantitative description of the experimental data. 

\begin{figure}
\includegraphics{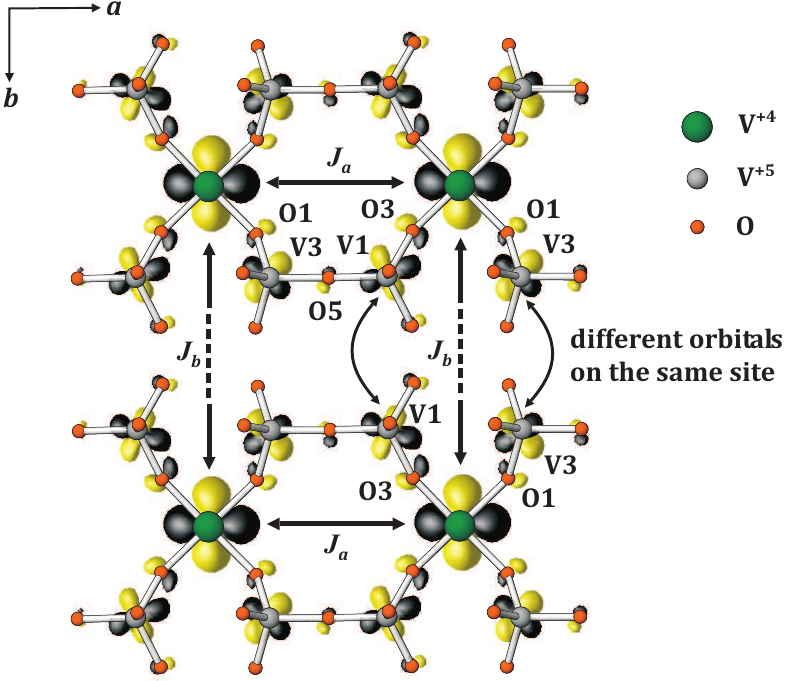}
\caption{\label{fig:wannier}
(Color online) Wannier functions based on the $d_{xy}$ orbitals of V$^{+4}$, and superexchange pathways for leading interactions $J_a$ and $J_b$. The pathways for $J_b$ are broken for clarity (compare to Figs.~\ref{fig:str} and~\ref{fig:dm}). The long-range coupling $J_a$ follows the V$^{+4}$--O1--V3--O5--V1--O3--V$^{+4}$ pathway and takes advantage of the sizable contribution of the O5 orbitals to the Wannier function. The coupling $J_b$ features shorter V$^{+4}$--O1--V3--O1--V$^{+4}$ and V$^{+4}$--O3--V1--O3--V$^{+4}$ pathways that, however, are relatively inefficient because of the $d$ orbitals of different symmetry overlapping on the V1 and V3 sites.
}
\end{figure}
The GGA+$U$ calculations support the results of our model analysis (Table~\ref{tab:exchange}). However, both $J_a$ and $J_b$ are strongly reduced with respect to their LDA-based $J_i^{\AFM}$ values. The relevant ferromagnetic (FM) contributions can be understood as a result of virtual hoppings from the half-filled $d_{xy}$ to the empty $d_{x^2-y^2}$ and $d_{3z^2-r^2}$ orbitals.\cite{mazurenko2006} These hoppings support the FM spin arrangement according to Hund's coupling in the V $3d$ shell. However, a quantitative description of this effect remains challenging, because the $d_{x^2-y^2}$ and $d_{3z^2-r^2}$ states are heavily mixed with those of V$^{+5}$.

It is quite difficult to make a link between our model and the earlier empirical $J_1\!-\!J_2$ model by Chakrabarty~\textit{et~al}.\cite{chakrabarty} The interaction $J_a$ is a nearest-neighbor coupling. However, it is likewise a long-range coupling connecting two V$^{+4}$ ions that are separated by 7.43~\r A. One could hardly envisage any further, next-nearest-neighbor coupling along $a$ that would be strong enough to induce any appreciable frustration. Indeed, we do not find such a coupling in our microscopic analysis. The $J_1-J_2$ chains can be probably formed along other directions (either $b$ or $c$), but the respective interactions are quite weak and render the ensuing frustration a secondary effect compared to the much larger, non-frustrated coupling $J_a$. Therefore, we conclude that the spin system of BaV$_3$O$_8$ is quasi-one-dimensional (1D) and only weakly frustrated. The frustration may originate from weak interchain interactions, but, as we will show below, it has no visible effect on the magnetic ordering temperature~$T_N$.

Our microscopic results are rather unexpected and contrast with the earlier knowledge on the superexchange in systems with non-magnetic polyanions. The shorter coupling $J_b$ runs via a double bridge of the V$^{+5}$O$_4$ tetrahedra (Fig.~\ref{fig:str}, left). Superexchange pathways of this type are rather efficient in phosphates, where properly aligned double bridges of PO$_4$ tetrahedra mediate sizable couplings, with $J$ reaching $100-150$~K.\cite{garrett1997,*tennant1997,nath2008} However, in BaV$_3$O$_8$ a similar pathway generates a relatively weak coupling $J_b$ that is exceeded by $J_a$, even though $J_a$ entails a much longer V--V pathway with two contiguous VO$_4$ tetrahedra (Fig.~\ref{fig:str}, left). The pathways of two contiguous non-magnetic tetrahedra are generally considered inactive in the superexchange. They do not induce any sizable and experimentally relevant couplings in the majority of Cu$^{+2}$ and V$^{+4}$ phosphates.\cite{tsirlin2008,janson2011} The only measurable coupling of this type is perhaps the interdimer exchange in BaCuSi$_2$O$_6$. It amounts to $7-8$~K according to the inelastic neutron scattering data\cite{ruegg2007} and DFT calculations.\cite{bacusi2o6}

The nature of the superexchange couplings is elucidated by the Wannier functions based on the $d_{xy}$ orbitals of V$^{+4}$ (Fig.~\ref{fig:wannier}). In addition to this central orbital, each Wannier function entails $2p$ orbitals of nearest-neighbor oxygen atoms (O1 and O3), $d$ orbitals of V$^{+5}$ (V1 and V3), and even the $2p$ orbitals of the third-neighbor O5 atoms that bridge two V$^{+5}$O$_4$ tetrahedra of the [V$_2$O$_7$] group. The leading coupling $J_a$ follows the remarkably long superexchange pathway V$^{+4}$--O1--V3--O5--V1--O3--V$^{+4}$ with the V$^{+4}$--V$^{+4}$ distance of 7.43~\r A. The weaker coupling $J_b$ features a shorter distance of 5.55~\r A and relies on the V$^{+4}$--O1--V3--O1--V$^{+4}$ and V$^{+4}$--O3--V1--O3--V$^{+4}$ pathways. However, each of these pathways is relatively inefficient, because $d$ orbitals of different symmetry overlap on the V1 and V3 sites. 

The couplings $J_a$ and $J_b$ are very sensitive to the nature of the non-magnetic polyanion. We performed a simple test by calculating the band structure and evaluating $t_i$'s in a fictitious compound BaVOAs$_2$O$_7$ having the same crystal structure as BaV$_3$O$_8$ and As$^{+5}$ ions in the positions of V$^{+5}$. Both V$^{+5}$ and As$^{+5}$ feature similar ionic radii but very different electronic structure. The low-lying $3d$ states are present in the former and absent in the latter ion. Our calculations for the hypothetical compound BaVOAs$_2$O$_7$ yield $t_a=4$~meV and $t_b=-121$~meV that essentially change the microscopic regime compared to $t_a=82$~meV and $t_b=-42$~meV in BaV$_3$O$_8$ (BaVOV$_2$O$_7$). Now the spin chains are directed along $b$ according to the the shorter pathway for $J_b$, while the coupling along $a$ is basically eliminated.

The role of V$^{+5}$ can be clearly seen from the nature of the states in the vicinity of the Fermi level (Fig.~\ref{fig:dos}). The low-lying $3d$ states of V$^{+5}$ provide 23.6~\% of states compared to 12.5~\% states of the O $2p$ origin. By contrast, in the fictitious BaVOAs$_2$O$_7$, the contribution of O $2p$ increases to 23.8~\%, whereas that of As$^{+5}$ is 11.0~\%, only. The large contribution of the V$^{+5}$ ions is also well seen in the V$^{+4}$ $d_{xy}$-based Wannier functions (Fig.~\ref{fig:wannier}).  

\begin{figure}
\includegraphics{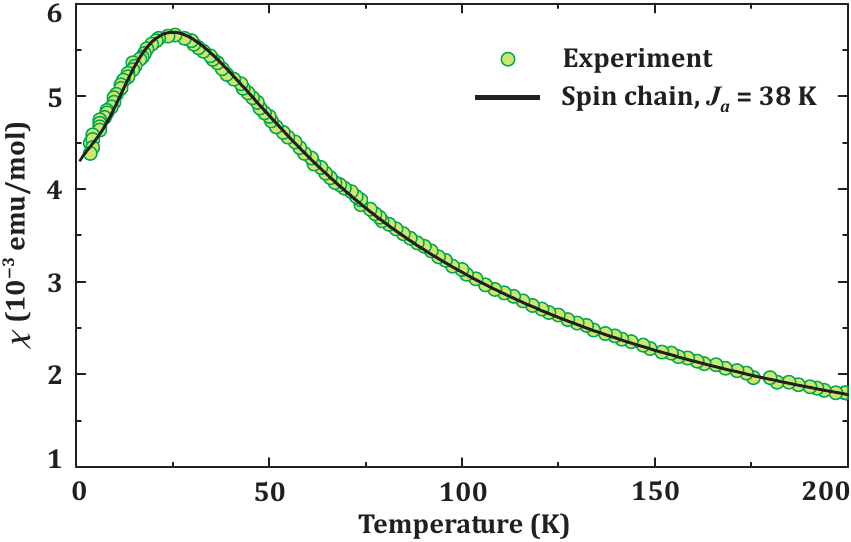}
\caption{\label{fig:chi}
(Color online) Magnetic susceptibility of BaV$_3$O$_8$ and its fit with the spin-chain model, as described in the text. Experimental data are from Ref.~\onlinecite{chakrabarty}.
}
\end{figure}

To compare our microscopic magnetic model with the experimental data, we first note that the spin lattice of BaV$_3$O$_8$ is quasi-1D, so that its thermodynamic properties should be well described by a well-known model of a uniform spin-$\frac12$ chain. Indeed, the experimental magnetic susceptibility can be fitted with the analytical expression for the uniform spin chain\cite{johnston2000} and yields $J_a\simeq 39$~K, $g\simeq 1.99$, $\chi_0\simeq 1.1\times 10^{-4}$~emu/mol (Fig.~\ref{fig:chi}). The resulting value of $J_a$ is in excellent agreement with DFT (Table~\ref{tab:exchange}), whereas the $g$-value slightly below 2.0 is quite typical for V$^{+4}$ compounds.\cite{tsirlin2011c,ivanshin2003,foerster2013} Similar to Ref.~\onlinecite{chakrabarty}, our fit does not extend below $10-15$\,K, where interchain couplings and the proximity to the AFM transition lead to deviations of the susceptibility from the 1D model. Nevertheless, our model should be preferred even on phenomenological grounds, because we successfully describe the susceptibility data using only one $J$ parameter, compared to Ref.~\onlinecite{chakrabarty}, where three parameters ($J_1$, $J_2$, and an interchain coupling) have been used. Moreover, the uniform-chain model is well-justified microscopically.

The long-range magnetic order is largely determined by the interchain couplings $J_b$, $J_{\perp1}$, and $J_{\perp2}$. However, a precise QMC simulation for this coupling regime is impossible because of the weak frustration pertaining to $J_{\perp1}$ and $J_{\perp2}$.\footnote{Each [VOV$_2$O$_7$] layer is shifted for $b/2$ with respect to its neighboring layers. Therefore, each V$^{+4}$ ion interacts with two V$^{+4}$ ions from the neighboring layer, which then leads to a triangle of frustrated AFM interactions $J_{\perp\alpha}-J_b-J_{\perp\alpha}$ ($\alpha=1,2$). While on the classical level this frustration would prevent the system from any 3D magnetic order, quantum fluctuations and/or magnetic anisotropy will typically induce the long-range order at low temperatures. Considering the low energies of interlayer couplings and relevant anisotropies, this coupling regime is quite challenging for a quantitative microscopic modeling.} We consider a simplified model and assume uniform, non-frustrated couplings $J_{\perp}^{\eff}$ that connect the layers along the $c$ direction (same couplings within the bilayer and between the bilayers). We deliberately choose $J_{\perp}^{\eff}=1$~K, which is the upper estimate of the interlayer exchange (Table~\ref{tab:exchange}), and show that even this regime of an overly strong interlayer coupling leads to a realistic estimate of $T_N$. This way, we conclude that the properties of BaV$_3$O$_8$ are not influenced by the frustration.

\begin{figure}
\includegraphics{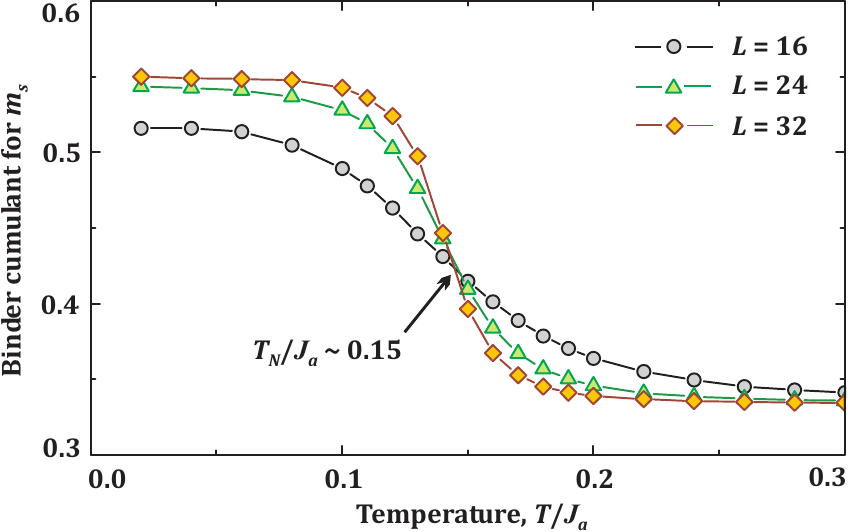}
\caption{\label{fig:binder}
(Color online) Binder cumulant for staggered magnetization ($m_s$) calculated for finite lattices of different size $L$. The model entails weakly coupled spin chains with $J_b/J_a=0.17$ and $J_{\perp}^{\eff}/J_a=0.025$. The crossing point is the N\'eel temperature $T_N/J_a\simeq 0.15$, i.e., $T_N\simeq 6$~K.
}
\end{figure}

We set $J_b/J_a=0.17$ and $J_{\perp}^{\eff}/J_a=0.025$ and perform QMC simulations for the relevant three-dimensional spin lattice. The N\'eel temperature is obtained as the crossing point of $B(T)$ curves, where $B(T)=\langle m_s^4\rangle/\langle m_s^2\rangle^2$ is Binder's cumulant for staggered magnetization (Fig.~\ref{fig:binder}).\cite{binder1997} Universal scaling properties ensure that $B(T_N)$ does not depend on the cluster size $L$ at $T_N$. Therefore, we get a precise estimate of $T_N/J_a\simeq 0.15$ as the crossing point of the $B(T)$ curves calculated for different $L$. We find $T_N\simeq 6$~K in remarkable agreement with the experiment ($T_N=5.5-6.0$~K).\cite{chakrabarty} 

\begin{figure}
\includegraphics{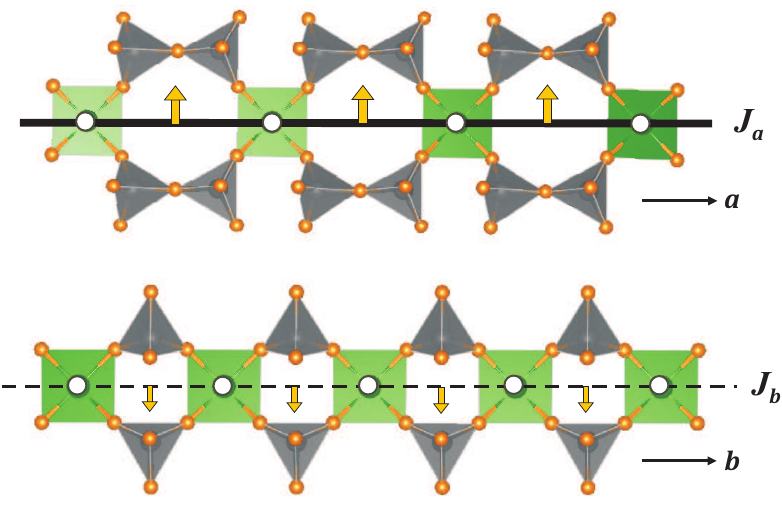}
\caption{\label{fig:dm}
(Color online) The couplings $J_a$ (top) and $J_b$ (bottom) and their relevant DM vectors shown by arrows (bond directions follow crystallographic directions). The vector length is schematic and does not represent the values of DM anisotropy in BaV$_3$O$_8$.
}
\end{figure}

Considering the high accuracy of our DFT results, we can go one step further and analyze the magnetic anisotropy of BaV$_3$O$_8$. Inversion centers of the $P2_1/m$ space group are located between the [VOV$_2$O$_7$] layers (Fig.~\ref{fig:str}, right). Therefore, individual layers lack inversion symmetry, so that the DM interactions are allowed for both $J_a$ and $J_b$. Moreover, the lattice periodicity along both $a$ and $b$ (Fig.~\ref{fig:str}, left) requires that every bond of $J_a$ or $J_b$ type features exactly the same DM vector (Fig.~\ref{fig:dm}). This is very different from the majority of anisotropic spin chains, where the unit cell includes two or more chain segments, the neighboring segments are related by a screw-axis or glide-plane symmetry, and a staggered (alternating) arrangement of the DM vectors is observed.\cite{[{For example: }][{}]affleck1999,*asano2000}

The mirror plane that is perpendicular to the $b$ axis and contains all V$^{+4}$ ions of the same spin chain implies that both $x$ and $z$ components of $\Dv_a$ are zero by symmetry (Fig.~\ref{fig:dm}).\cite{moriya1960} However, its $y$-component is non-zero, and indeed we find $D_a^y\simeq 0.1$~K, while $D_a^x$ and $D_a^z$ are below 0.01\,K and, thus, equal to zero within the accuracy of our computational method.\footnote{%
Magnetic anisotropy parameters are extracted from total energies, which are converged to $10^{-6}$\,eV/f.u. However, numerical issues, including the $k$-point integration, lead to a somewhat lower accuracy that we estimate as $10^{-5}$\,eV/f.u. (i.e., about 0.05\,K for individual magnetic couplings) according to the mismatch between total energies of same spin configurations calculated in different supercells.} The DM vector $\Dv_b$ should lie in the $ac$ plane, but its value is also well below the accuracy of our method, because even the isotropic exchange $J_b$ is quite weak. Nevertheless, an important and robust finding is that BaV$_3$O$_8$ features at least a weak DM anisotropy $|\Dv_a|/J_a\simeq 0.003$, which is allowed by symmetry, and this anisotropy is uniform along the $a$ and $b$ directions. In the following, we will discuss possible implications of this result for other vanadate compounds.     

\section{Discussion and Summary}  
We have shown that the magnetism of BaV$_3$O$_8$ is well described by a model of weakly coupled non-frustrated spin chains running along the crystallographic $a$ direction. The coupling regime of this compound is largely determined by the V$^{+5}$ ions with low-lying $3d$ orbitals that contribute to the ``magnetic orbitals'' ($d_{xy}$-type Wannier function centered on V$^{+4}$, Fig.~\ref{fig:wannier}) and influence the electron hoppings. The effect of ions centering non-magnetic tetrahedra is well-known from earlier studies.\cite{janson2011} However, the case of BaV$_3$O$_8$ is somewhat special, because V$^{+5}$ eliminates the anticipated coupling along $b$ (shorter V$^{+4}$--V$^{+4}$ pathway, double bridge of the VO$_4$ tetrahedra: Fig.~\ref{fig:dm}, bottom) and triggers the unanticipated coupling along $a$ (longer V$^{+4}$--V$^{+4}$ pathway, a bridge of two contiguous VO$_4$ tetrahedra: Fig.~\ref{fig:dm}, top). This microscopic regime calls for a further analysis of the magnetic exchange in vanadate compounds. 

Many of the Cu$^{+2}$ and V$^{+4}$ vanadates show the coupling regime, which is at least qualitatively similar to that of phosphates, arsenates and other compounds with non-magnetic polyanions that do not feature low-lying empty $d$ orbitals. Double bridges of the V$^{+5}$O$_4$ tetrahedra typically lead to a sizable AFM coupling, as in $\beta$-Cu$_2$V$_2$O$_7$ ($J\simeq 60$~K)\cite{tsirlin2010} and CsV$_2$O$_5$ ($J\simeq 90$~K).\cite{saul2011} Single bridges of the V$^{+5}$O$_4$ tetrahedra lead to a weaker exchange, as in Pb$_2$V$_3$O$_9$ ($J_1\simeq 31$~K, $J_1'\simeq 20$~K).\cite{tsirlin2011b} However, this trend is broken in a structurally related compound Sr$_2$V$_3$O$_9$ with $J\simeq 82$~K running through a single VO$_4$ tetrahedron.\cite{kaul2003} BaV$_3$O$_8$ shows a further departure from this general trend, because the double bridges become very inefficient ($J_b\simeq 6$~K), but longer pathways of two contiguous tetrahedra play an important role instead.

We argue that not only the nature of the bridge and not only the number of the bridging tetrahedra, but also subtle details of their mutual orientation play decisive role for the superexchange in vanadates. In contrast with $\beta$-Cu$_2$V$_2$O$_7$ and CsV$_2$O$_5$, the two VO$_4$ tetrahedra of the double bridge ($J_b$) in BaV$_3$O$_8$ are not related by the inversion symmetry. They are found on one side of the chain (Fig.~\ref{fig:str}, right and Fig.~\ref{fig:dm}, bottom), which is what probably makes the microscopic scenario of BaV$_3$O$_8$ so different from that of the previously known compounds. Two other examples should be mentioned as well. BaAg$_2$Cu(VO$_4)_2$ entails spin chains with the magnetic CuO$_4$ units bridged by VO$_4$ tetrahedra. Two slightly nonequivalent structural chains are drastically different in terms of their magnetism, because tiny differences in the positions of the VO$_4$ tetrahedra render the coupling either FM or AFM.\cite{tsirlin2012} This example demonstrates the crucial role of structural details in vanadate compounds. Yet another example is given by the ACuV$_2$O$_7$ compounds (A = Sr, Ba) that are remarkably similar to BaV$_3$O$_8$ and feature same structure of the layer (Fig.~\ref{fig:str}, left), with the VO$_5$ square pyramids replaced by the CuO$_4$ planar units. Leading magnetic couplings in ACuV$_2$O$_7$ are FM,\cite{belik2005} although the long-range nature of these couplings should induce AFM exchange. While details of the magnetic model of ACuV$_2$O$_7$ are not known yet, their microscopic analysis could be insightful.

An interesting aspect of the vanadate compounds is their ability to form spin chains with the uniform DM anisotropy. The presence of the DM anisotropy and the mutual orientation of the DM vectors are determined by different aspects of the crystal structures and chemical composition. While these features are hard to track systematically, it is remarkable that vanadates and only vanadates feature relevant structural details: i) the lack of the inversion symmetry between the magnetic ions; ii) one chain segment per crystallographic unit cell. The latter feature implies that neighboring segments of the chain are related by the translation symmetry and host exactly the same DM vectors, in contrast to a more conventional regime of alternating DM vectors (staggered DM anisotropy). 

The regime of the uniform DM anisotropy on a spin chain hitherto lacks any material prototypes, while theory supplies intriguing predictions\cite{starykh2008,garate2010} that still await their experimental verification. The weak DM anisotropy should trigger a weak helical modulation along the chain direction, whereas a stronger anisotropy leads to a dimerized AFM phase.\cite{garate2010} These phases could be probed by neutron-scattering experiments. Moreover, interesting predictions regarding the line splitting in electron spin resonance spectra are available.\cite{starykh2008} 

Although BaV$_3$O$_8$ features only a weak DM anisotropy, it gives ideas for a deliberate design of materials with larger DM vectors uniformly arranged on a spin-$\frac12$ chain. For example, one could switch the chain direction and exploit the DM couplings generated by double bridges of the VO$_4$ tetrahedra ($J_b$), as opposed to the bridges of two contiguous tetrahedra in the $J_a$ coupling in BaV$_3$O$_8$. The chain direction would be switched in the hypothetical compound BaVOAs$_2$O$_7$ or, even more conveniently, in the existing compound Na$_2$SrV$_3$O$_9$, where V$^{+4}$O$_5$ pyramids are bridged by V$^{+5}$O$_4$ tetrahedra and form isolated chains.\cite{shpanchenko2003} These chains can be envisaged by cleaving the [VOV$_2$O$_7$] layers at the bridging O5 oxygens of the V$_2$O$_7$ groups. The sizable intrachain coupling of $J\simeq 80$~K in Na$_2$SrV$_3$O$_9$ makes it an interesting compound for future experimental and microscopic studies.\cite{shpanchenko2003} Similar effects should be expected in the structurally related ACuV$_2$O$_7$ with A = Sr and Ba,\cite{belik2005} although the direction of spin chains in these compounds is yet to be determined.

In summary, we have shown that the spin-$\frac12$ magnet BaV$_3$O$_8$ is well described by a model of weakly coupled spin chains with the intrachain exchange $J_a\simeq 38$~K. The N\'eel temperature of $T_N\simeq 6$~K is reproduced even in a non-frustrated spin model assuming the strongest possible interchain coupling. The combination of the magnetic V$^{+4}$ and non-magnetic V$^{+5}$ ions causes the admixture of the low-lying but empty $3d$ states of V$^{+5}$ to the half-filled states of V$^{+4}$, thus triggering an unusual coupling regime. Remarkably, the spin chains are directed along $a$, even though the respective coupling $J_a$ features a very long superexchange pathway that was previously deemed unfavorable for the superexchange. The shorter pathway along $b$ results in a weaker coupling. We argue that BaV$_3$O$_8$ is a rare compound showing the uniform arrangement of the DM vectors along the spin chains. While the DM anisotropy in BaV$_3$O$_8$ is quite weak, general trends evidenced by this compound suggest future directions for the search of systems with uniform DM anisotropy.      

\acknowledgments
The author acknowledges fruitful discussions with Oleg Janson and Vladimir Mazurenko, and the financial support from the ESF through the Mobilitas Grant MTT77.


\begin{thebibliography}{54}%
\makeatletter
\providecommand \@ifxundefined [1]{%
 \@ifx{#1\undefined}
}%
\providecommand \@ifnum [1]{%
 \ifnum #1\expandafter \@firstoftwo
 \else \expandafter \@secondoftwo
 \fi
}%
\providecommand \@ifx [1]{%
 \ifx #1\expandafter \@firstoftwo
 \else \expandafter \@secondoftwo
 \fi
}%
\providecommand \natexlab [1]{#1}%
\providecommand \enquote  [1]{``#1''}%
\providecommand \bibnamefont  [1]{#1}%
\providecommand \bibfnamefont [1]{#1}%
\providecommand \citenamefont [1]{#1}%
\providecommand \href@noop [0]{\@secondoftwo}%
\providecommand \href [0]{\begingroup \@sanitize@url \@href}%
\providecommand \@href[1]{\@@startlink{#1}\@@href}%
\providecommand \@@href[1]{\endgroup#1\@@endlink}%
\providecommand \@sanitize@url [0]{\catcode `\\12\catcode `\$12\catcode
  `\&12\catcode `\#12\catcode `\^12\catcode `\_12\catcode `\%12\relax}%
\providecommand \@@startlink[1]{}%
\providecommand \@@endlink[0]{}%
\providecommand \url  [0]{\begingroup\@sanitize@url \@url }%
\providecommand \@url [1]{\endgroup\@href {#1}{\urlprefix }}%
\providecommand \urlprefix  [0]{URL }%
\providecommand \Eprint [0]{\href }%
\providecommand \doibase [0]{http://dx.doi.org/}%
\providecommand \selectlanguage [0]{\@gobble}%
\providecommand \bibinfo  [0]{\@secondoftwo}%
\providecommand \bibfield  [0]{\@secondoftwo}%
\providecommand \translation [1]{[#1]}%
\providecommand \BibitemOpen [0]{}%
\providecommand \bibitemStop [0]{}%
\providecommand \bibitemNoStop [0]{.\EOS\space}%
\providecommand \EOS [0]{\spacefactor3000\relax}%
\providecommand \BibitemShut  [1]{\csname bibitem#1\endcsname}%
\let\auto@bib@innerbib\@empty
\bibitem [{\citenamefont {Balents}(2010)}]{balents2010}%
  \BibitemOpen
  \bibfield  {author} {\bibinfo {author} {\bibfnamefont {L.}~\bibnamefont
  {Balents}},\ }\href@noop {} {\bibfield  {journal} {\bibinfo  {journal}
  {Nature}\ }\textbf {\bibinfo {volume} {464}},\ \bibinfo {pages} {199}
  (\bibinfo {year} {2010})}\BibitemShut {NoStop}%
\bibitem [{\citenamefont {Han}\ \emph {et~al.}(2012)\citenamefont {Han},
  \citenamefont {Helton}, \citenamefont {Chu}, \citenamefont {Nocera},
  \citenamefont {Rodriguez-Rivera}, \citenamefont {Broholm},\ and\
  \citenamefont {Lee}}]{han2012}%
  \BibitemOpen
  \bibfield  {author} {\bibinfo {author} {\bibfnamefont {T.-H.}\ \bibnamefont
  {Han}}, \bibinfo {author} {\bibfnamefont {J.~S.}\ \bibnamefont {Helton}},
  \bibinfo {author} {\bibfnamefont {S.}~\bibnamefont {Chu}}, \bibinfo {author}
  {\bibfnamefont {D.~G.}\ \bibnamefont {Nocera}}, \bibinfo {author}
  {\bibfnamefont {J.~A.}\ \bibnamefont {Rodriguez-Rivera}}, \bibinfo {author}
  {\bibfnamefont {C.}~\bibnamefont {Broholm}}, \ and\ \bibinfo {author}
  {\bibfnamefont {Y.~S.}\ \bibnamefont {Lee}},\ }\href@noop {} {\bibfield
  {journal} {\bibinfo  {journal} {Nature}\ }\textbf {\bibinfo {volume} {492}},\
  \bibinfo {pages} {406} (\bibinfo {year} {2012})}\BibitemShut {NoStop}%
\bibitem [{\citenamefont {Greedan}(2001)}]{greedan2001}%
  \BibitemOpen
  \bibfield  {author} {\bibinfo {author} {\bibfnamefont {J.}~\bibnamefont
  {Greedan}},\ }\href@noop {} {\bibfield  {journal} {\bibinfo  {journal} {J.
  Mater. Chem}\ }\textbf {\bibinfo {volume} {11}},\ \bibinfo {pages} {37}
  (\bibinfo {year} {2001})}\BibitemShut {NoStop}%
\bibitem [{\citenamefont {Yasuda}\ \emph {et~al.}(2005)\citenamefont {Yasuda},
  \citenamefont {Todo}, \citenamefont {Hukushima}, \citenamefont {Alet},
  \citenamefont {Keller}, \citenamefont {Troyer},\ and\ \citenamefont
  {Takayama}}]{yasuda2005}%
  \BibitemOpen
  \bibfield  {author} {\bibinfo {author} {\bibfnamefont {C.}~\bibnamefont
  {Yasuda}}, \bibinfo {author} {\bibfnamefont {S.}~\bibnamefont {Todo}},
  \bibinfo {author} {\bibfnamefont {K.}~\bibnamefont {Hukushima}}, \bibinfo
  {author} {\bibfnamefont {F.}~\bibnamefont {Alet}}, \bibinfo {author}
  {\bibfnamefont {M.}~\bibnamefont {Keller}}, \bibinfo {author} {\bibfnamefont
  {M.}~\bibnamefont {Troyer}}, \ and\ \bibinfo {author} {\bibfnamefont
  {H.}~\bibnamefont {Takayama}},\ }\href@noop {} {\bibfield  {journal}
  {\bibinfo  {journal} {Phys. Rev. Lett.}\ }\textbf {\bibinfo {volume} {94}},\
  \bibinfo {pages} {217201} (\bibinfo {year} {2005})}\BibitemShut {NoStop}%
\bibitem [{\citenamefont {Momma}\ and\ \citenamefont {Izumi}(2011)}]{vesta}%
  \BibitemOpen
  \bibfield  {author} {\bibinfo {author} {\bibfnamefont {K.}~\bibnamefont
  {Momma}}\ and\ \bibinfo {author} {\bibfnamefont {F.}~\bibnamefont {Izumi}},\
  }\href@noop {} {\bibfield  {journal} {\bibinfo  {journal} {J. Appl.
  Crystallogr.}\ }\textbf {\bibinfo {volume} {44}},\ \bibinfo {pages} {1272}
  (\bibinfo {year} {2011})}\BibitemShut {NoStop}%
\bibitem [{\citenamefont {Chakrabarty}\ \emph {et~al.}(2013)\citenamefont
  {Chakrabarty}, \citenamefont {Mahajan}, \citenamefont {Gippius},
  \citenamefont {Tkachev}, \citenamefont {B\"uttgen},\ and\ \citenamefont
  {Kraetschmer}}]{chakrabarty}%
  \BibitemOpen
  \bibfield  {author} {\bibinfo {author} {\bibfnamefont {T.}~\bibnamefont
  {Chakrabarty}}, \bibinfo {author} {\bibfnamefont {A.~V.}\ \bibnamefont
  {Mahajan}}, \bibinfo {author} {\bibfnamefont {A.~A.}\ \bibnamefont
  {Gippius}}, \bibinfo {author} {\bibfnamefont {A.~V.}\ \bibnamefont
  {Tkachev}}, \bibinfo {author} {\bibfnamefont {N.}~\bibnamefont {B\"uttgen}},
  \ and\ \bibinfo {author} {\bibfnamefont {W.}~\bibnamefont {Kraetschmer}},\
  }\href@noop {} {\bibfield  {journal} {\bibinfo  {journal} {Phys. Rev. B}\
  }\textbf {\bibinfo {volume} {88}},\ \bibinfo {pages} {014433} (\bibinfo
  {year} {2013})}\BibitemShut {NoStop}%
\bibitem [{\citenamefont {Gangadharaiah}\ \emph {et~al.}(2008)\citenamefont
  {Gangadharaiah}, \citenamefont {Sun},\ and\ \citenamefont
  {Starykh}}]{starykh2008}%
  \BibitemOpen
  \bibfield  {author} {\bibinfo {author} {\bibfnamefont {S.}~\bibnamefont
  {Gangadharaiah}}, \bibinfo {author} {\bibfnamefont {J.}~\bibnamefont {Sun}},
  \ and\ \bibinfo {author} {\bibfnamefont {O.~A.}\ \bibnamefont {Starykh}},\
  }\href@noop {} {\bibfield  {journal} {\bibinfo  {journal} {Phys. Rev. B}\
  }\textbf {\bibinfo {volume} {78}},\ \bibinfo {pages} {054436} (\bibinfo
  {year} {2008})}\BibitemShut {NoStop}%
\bibitem [{\citenamefont {Garate}\ and\ \citenamefont
  {Affleck}(2010)}]{garate2010}%
  \BibitemOpen
  \bibfield  {author} {\bibinfo {author} {\bibfnamefont {I.}~\bibnamefont
  {Garate}}\ and\ \bibinfo {author} {\bibfnamefont {I.}~\bibnamefont
  {Affleck}},\ }\href@noop {} {\bibfield  {journal} {\bibinfo  {journal} {Phys.
  Rev. B}\ }\textbf {\bibinfo {volume} {81}},\ \bibinfo {pages} {144419}
  (\bibinfo {year} {2010})}\BibitemShut {NoStop}%
\bibitem [{\citenamefont {Oka}\ \emph {et~al.}(1995)\citenamefont {Oka},
  \citenamefont {Yao},\ and\ \citenamefont {Yamamoto}}]{oka1995}%
  \BibitemOpen
  \bibfield  {author} {\bibinfo {author} {\bibfnamefont {Y.}~\bibnamefont
  {Oka}}, \bibinfo {author} {\bibfnamefont {T.}~\bibnamefont {Yao}}, \ and\
  \bibinfo {author} {\bibfnamefont {N.}~\bibnamefont {Yamamoto}},\ }\href@noop
  {} {\bibfield  {journal} {\bibinfo  {journal} {J. Solid State Chem.}\
  }\textbf {\bibinfo {volume} {117}},\ \bibinfo {pages} {407} (\bibinfo {year}
  {1995})}\BibitemShut {NoStop}%
\bibitem [{\citenamefont {Marsh}(1996)}]{marsh1996}%
  \BibitemOpen
  \bibfield  {author} {\bibinfo {author} {\bibfnamefont {R.~E.}\ \bibnamefont
  {Marsh}},\ }\href@noop {} {\bibfield  {journal} {\bibinfo  {journal} {J.
  Solid State Chem.}\ }\textbf {\bibinfo {volume} {122}},\ \bibinfo {pages}
  {245} (\bibinfo {year} {1996})}\BibitemShut {NoStop}%
\bibitem [{\citenamefont {Kaul}\ \emph {et~al.}(2003)\citenamefont {Kaul},
  \citenamefont {Rosner}, \citenamefont {Yushankhai}, \citenamefont
  {Sichelschmidt}, \citenamefont {Shpanchenko},\ and\ \citenamefont
  {Geibel}}]{kaul2003}%
  \BibitemOpen
  \bibfield  {author} {\bibinfo {author} {\bibfnamefont {E.~E.}\ \bibnamefont
  {Kaul}}, \bibinfo {author} {\bibfnamefont {H.}~\bibnamefont {Rosner}},
  \bibinfo {author} {\bibfnamefont {V.}~\bibnamefont {Yushankhai}}, \bibinfo
  {author} {\bibfnamefont {J.}~\bibnamefont {Sichelschmidt}}, \bibinfo {author}
  {\bibfnamefont {R.~V.}\ \bibnamefont {Shpanchenko}}, \ and\ \bibinfo {author}
  {\bibfnamefont {C.}~\bibnamefont {Geibel}},\ }\href@noop {} {\bibfield
  {journal} {\bibinfo  {journal} {Phys. Rev. B}\ }\textbf {\bibinfo {volume}
  {67}},\ \bibinfo {pages} {174417} (\bibinfo {year} {2003})}\BibitemShut
  {NoStop}%
\bibitem [{\citenamefont {Tsirlin}\ \emph {et~al.}(2008)\citenamefont
  {Tsirlin}, \citenamefont {Nath}, \citenamefont {Geibel},\ and\ \citenamefont
  {Rosner}}]{tsirlin2008}%
  \BibitemOpen
  \bibfield  {author} {\bibinfo {author} {\bibfnamefont {A.~A.}\ \bibnamefont
  {Tsirlin}}, \bibinfo {author} {\bibfnamefont {R.}~\bibnamefont {Nath}},
  \bibinfo {author} {\bibfnamefont {C.}~\bibnamefont {Geibel}}, \ and\ \bibinfo
  {author} {\bibfnamefont {H.}~\bibnamefont {Rosner}},\ }\href@noop {}
  {\bibfield  {journal} {\bibinfo  {journal} {Phys. Rev. B}\ }\textbf {\bibinfo
  {volume} {77}},\ \bibinfo {pages} {104436} (\bibinfo {year}
  {2008})}\BibitemShut {NoStop}%
\bibitem [{\citenamefont {Perdew}\ and\ \citenamefont {Wang}(1992)}]{pw92}%
  \BibitemOpen
  \bibfield  {author} {\bibinfo {author} {\bibfnamefont {J.~P.}\ \bibnamefont
  {Perdew}}\ and\ \bibinfo {author} {\bibfnamefont {Y.}~\bibnamefont {Wang}},\
  }\href@noop {} {\bibfield  {journal} {\bibinfo  {journal} {Phys. Rev. B}\
  }\textbf {\bibinfo {volume} {45}},\ \bibinfo {pages} {13244} (\bibinfo {year}
  {1992})}\BibitemShut {NoStop}%
\bibitem [{\citenamefont {Perdew}\ \emph {et~al.}(1996)\citenamefont {Perdew},
  \citenamefont {Burke},\ and\ \citenamefont {Ernzerhof}}]{pbe96}%
  \BibitemOpen
  \bibfield  {author} {\bibinfo {author} {\bibfnamefont {J.~P.}\ \bibnamefont
  {Perdew}}, \bibinfo {author} {\bibfnamefont {K.}~\bibnamefont {Burke}}, \
  and\ \bibinfo {author} {\bibfnamefont {M.}~\bibnamefont {Ernzerhof}},\
  }\href@noop {} {\bibfield  {journal} {\bibinfo  {journal} {Phys. Rev. Lett.}\
  }\textbf {\bibinfo {volume} {77}},\ \bibinfo {pages} {3865} (\bibinfo {year}
  {1996})}\BibitemShut {NoStop}%
\bibitem [{\citenamefont {Koepernik}\ and\ \citenamefont
  {Eschrig}(1999)}]{fplo}%
  \BibitemOpen
  \bibfield  {author} {\bibinfo {author} {\bibfnamefont {K.}~\bibnamefont
  {Koepernik}}\ and\ \bibinfo {author} {\bibfnamefont {H.}~\bibnamefont
  {Eschrig}},\ }\href@noop {} {\bibfield  {journal} {\bibinfo  {journal} {Phys.
  Rev. B}\ }\textbf {\bibinfo {volume} {59}},\ \bibinfo {pages} {1743}
  (\bibinfo {year} {1999})}\BibitemShut {NoStop}%
\bibitem [{\citenamefont {Kresse}\ and\ \citenamefont
  {Furthm\"uller}(1996{\natexlab{a}})}]{vasp1}%
  \BibitemOpen
  \bibfield  {author} {\bibinfo {author} {\bibfnamefont {G.}~\bibnamefont
  {Kresse}}\ and\ \bibinfo {author} {\bibfnamefont {J.}~\bibnamefont
  {Furthm\"uller}},\ }\href@noop {} {\bibfield  {journal} {\bibinfo  {journal}
  {Comput. Mater. Sci.}\ }\textbf {\bibinfo {volume} {6}},\ \bibinfo {pages}
  {15} (\bibinfo {year} {1996}{\natexlab{a}})}\BibitemShut {NoStop}%
\bibitem [{\citenamefont {Kresse}\ and\ \citenamefont
  {Furthm\"uller}(1996{\natexlab{b}})}]{vasp2}%
  \BibitemOpen
  \bibfield  {author} {\bibinfo {author} {\bibfnamefont {G.}~\bibnamefont
  {Kresse}}\ and\ \bibinfo {author} {\bibfnamefont {J.}~\bibnamefont
  {Furthm\"uller}},\ }\href@noop {} {\bibfield  {journal} {\bibinfo  {journal}
  {Phys. Rev. B}\ }\textbf {\bibinfo {volume} {54}},\ \bibinfo {pages} {11169}
  (\bibinfo {year} {1996}{\natexlab{b}})}\BibitemShut {NoStop}%
\bibitem [{\citenamefont {Tsirlin}\ and\ \citenamefont
  {Rosner}(2011)}]{tsirlin2011b}%
  \BibitemOpen
  \bibfield  {author} {\bibinfo {author} {\bibfnamefont {A.~A.}\ \bibnamefont
  {Tsirlin}}\ and\ \bibinfo {author} {\bibfnamefont {H.}~\bibnamefont
  {Rosner}},\ }\href@noop {} {\bibfield  {journal} {\bibinfo  {journal} {Phys.
  Rev. B}\ }\textbf {\bibinfo {volume} {83}},\ \bibinfo {pages} {064415}
  (\bibinfo {year} {2011})}\BibitemShut {NoStop}%
\bibitem [{\citenamefont {Tsirlin}\ \emph
  {et~al.}(2011{\natexlab{a}})\citenamefont {Tsirlin}, \citenamefont {Nath},
  \citenamefont {Sichelschmidt}, \citenamefont {Skourski}, \citenamefont
  {Geibel},\ and\ \citenamefont {Rosner}}]{tsirlin2011c}%
  \BibitemOpen
  \bibfield  {author} {\bibinfo {author} {\bibfnamefont {A.~A.}\ \bibnamefont
  {Tsirlin}}, \bibinfo {author} {\bibfnamefont {R.}~\bibnamefont {Nath}},
  \bibinfo {author} {\bibfnamefont {J.}~\bibnamefont {Sichelschmidt}}, \bibinfo
  {author} {\bibfnamefont {Y.}~\bibnamefont {Skourski}}, \bibinfo {author}
  {\bibfnamefont {C.}~\bibnamefont {Geibel}}, \ and\ \bibinfo {author}
  {\bibfnamefont {H.}~\bibnamefont {Rosner}},\ }\href@noop {} {\bibfield
  {journal} {\bibinfo  {journal} {Phys. Rev. B}\ }\textbf {\bibinfo {volume}
  {83}},\ \bibinfo {pages} {144412} (\bibinfo {year}
  {2011}{\natexlab{a}})}\BibitemShut {NoStop}%
\bibitem [{Note1()}]{Note1}%
  \BibitemOpen
  \bibinfo {note} {We use the fully-localized-limit double-counting correction,
  which is common to \protect \texttt {FPLO} and \protect \texttt
  {VASP}.}\BibitemShut {Stop}%
\bibitem [{\citenamefont {Korotin}\ \emph {et~al.}(1999)\citenamefont
  {Korotin}, \citenamefont {Elfimov}, \citenamefont {Anisimov}, \citenamefont
  {Troyer},\ and\ \citenamefont {Khomskii}}]{korotin1999}%
  \BibitemOpen
  \bibfield  {author} {\bibinfo {author} {\bibfnamefont {M.~A.}\ \bibnamefont
  {Korotin}}, \bibinfo {author} {\bibfnamefont {I.~S.}\ \bibnamefont
  {Elfimov}}, \bibinfo {author} {\bibfnamefont {V.~I.}\ \bibnamefont
  {Anisimov}}, \bibinfo {author} {\bibfnamefont {M.}~\bibnamefont {Troyer}}, \
  and\ \bibinfo {author} {\bibfnamefont {D.~I.}\ \bibnamefont {Khomskii}},\
  }\href@noop {} {\bibfield  {journal} {\bibinfo  {journal} {Phys. Rev. Lett.}\
  }\textbf {\bibinfo {volume} {83}},\ \bibinfo {pages} {1387} (\bibinfo {year}
  {1999})}\BibitemShut {NoStop}%
\bibitem [{\citenamefont {Korotin}\ \emph {et~al.}(2000)\citenamefont
  {Korotin}, \citenamefont {Anisimov}, \citenamefont {Saha-Dasgupta},\ and\
  \citenamefont {Dasgupta}}]{korotin2000}%
  \BibitemOpen
  \bibfield  {author} {\bibinfo {author} {\bibfnamefont {M.~A.}\ \bibnamefont
  {Korotin}}, \bibinfo {author} {\bibfnamefont {V.~I.}\ \bibnamefont
  {Anisimov}}, \bibinfo {author} {\bibfnamefont {T.}~\bibnamefont
  {Saha-Dasgupta}}, \ and\ \bibinfo {author} {\bibfnamefont {I.}~\bibnamefont
  {Dasgupta}},\ }\href@noop {} {\bibfield  {journal} {\bibinfo  {journal} {J.
  Phys.: Cond. Matter}\ }\textbf {\bibinfo {volume} {12}},\ \bibinfo {pages}
  {113} (\bibinfo {year} {2000})}\BibitemShut {NoStop}%
\bibitem [{\citenamefont {Tsirlin}\ \emph
  {et~al.}(2011{\natexlab{b}})\citenamefont {Tsirlin}, \citenamefont {Janson},\
  and\ \citenamefont {Rosner}}]{tsirlin2011}%
  \BibitemOpen
  \bibfield  {author} {\bibinfo {author} {\bibfnamefont {A.~A.}\ \bibnamefont
  {Tsirlin}}, \bibinfo {author} {\bibfnamefont {O.}~\bibnamefont {Janson}}, \
  and\ \bibinfo {author} {\bibfnamefont {H.}~\bibnamefont {Rosner}},\
  }\href@noop {} {\bibfield  {journal} {\bibinfo  {journal} {Phys. Rev. B}\
  }\textbf {\bibinfo {volume} {84}},\ \bibinfo {pages} {144429} (\bibinfo
  {year} {2011}{\natexlab{b}})}\BibitemShut {NoStop}%
\bibitem [{\citenamefont {Tsirlin}\ and\ \citenamefont
  {Rosner}(2009)}]{tsirlin2009}%
  \BibitemOpen
  \bibfield  {author} {\bibinfo {author} {\bibfnamefont {A.~A.}\ \bibnamefont
  {Tsirlin}}\ and\ \bibinfo {author} {\bibfnamefont {H.}~\bibnamefont
  {Rosner}},\ }\href@noop {} {\bibfield  {journal} {\bibinfo  {journal} {Phys.
  Rev. B}\ }\textbf {\bibinfo {volume} {79}},\ \bibinfo {pages} {214417}
  (\bibinfo {year} {2009})}\BibitemShut {NoStop}%
\bibitem [{\citenamefont {Xiang}\ \emph {et~al.}(2011)\citenamefont {Xiang},
  \citenamefont {Kan}, \citenamefont {Wei}, \citenamefont {Whangbo},\ and\
  \citenamefont {Gong}}]{xiang2011}%
  \BibitemOpen
  \bibfield  {author} {\bibinfo {author} {\bibfnamefont {H.~J.}\ \bibnamefont
  {Xiang}}, \bibinfo {author} {\bibfnamefont {E.~J.}\ \bibnamefont {Kan}},
  \bibinfo {author} {\bibfnamefont {S.-H.}\ \bibnamefont {Wei}}, \bibinfo
  {author} {\bibfnamefont {M.-H.}\ \bibnamefont {Whangbo}}, \ and\ \bibinfo
  {author} {\bibfnamefont {X.~G.}\ \bibnamefont {Gong}},\ }\href@noop {}
  {\bibfield  {journal} {\bibinfo  {journal} {Phys. Rev. B}\ }\textbf {\bibinfo
  {volume} {84}},\ \bibinfo {pages} {224429} (\bibinfo {year}
  {2011})}\BibitemShut {NoStop}%
\bibitem [{\citenamefont {Tsirlin}\ \emph {et~al.}()\citenamefont {Tsirlin},
  \citenamefont {Janson}, \citenamefont {Rousochatzakis},\ and\ \citenamefont
  {Rosner}}]{anisotropy}%
  \BibitemOpen
  \bibfield  {author} {\bibinfo {author} {\bibfnamefont {A.~A.}\ \bibnamefont
  {Tsirlin}}, \bibinfo {author} {\bibfnamefont {O.}~\bibnamefont {Janson}},
  \bibinfo {author} {\bibfnamefont {I.}~\bibnamefont {Rousochatzakis}}, \ and\
  \bibinfo {author} {\bibfnamefont {H.}~\bibnamefont {Rosner}},\ }\href@noop {}
  {}\bibinfo {note} {(unpublished)}\BibitemShut {NoStop}%
\bibitem [{sup()}]{supplement}%
  \BibitemOpen
  \href@noop {} {}\bibinfo {note} {See Supplemental Material for details of the
  computational procedure and an application of the four-configuration method
  to the evaluation of DM interactions. In particular, we demonstrate that all
  symmetry requirements for the DM vectors are reproduced
  computationally.}\BibitemShut {Stop}%
\bibitem [{\citenamefont {Todo}\ and\ \citenamefont {Kato}(2001)}]{loop}%
  \BibitemOpen
  \bibfield  {author} {\bibinfo {author} {\bibfnamefont {S.}~\bibnamefont
  {Todo}}\ and\ \bibinfo {author} {\bibfnamefont {K.}~\bibnamefont {Kato}},\
  }\href@noop {} {\bibfield  {journal} {\bibinfo  {journal} {Phys. Rev. Lett.}\
  }\textbf {\bibinfo {volume} {87}},\ \bibinfo {pages} {047203} (\bibinfo
  {year} {2001})}\BibitemShut {NoStop}%
\bibitem [{\citenamefont {Albuquerque}\ \emph {et~al.}(2007)\citenamefont
  {Albuquerque}, \citenamefont {Alet}, \citenamefont {Corboz}, \citenamefont
  {Dayal}, \citenamefont {Feiguin}, \citenamefont {Fuchs}, \citenamefont
  {Gamper}, \citenamefont {Gull}, \citenamefont {G\"urtler}, \citenamefont
  {Honecker}, \citenamefont {Igarashi}, \citenamefont {K\"orner}, \citenamefont
  {Kozhevnikov}, \citenamefont {L\"auchli}, \citenamefont {Manmana},
  \citenamefont {Matsumoto}, \citenamefont {McCulloch}, \citenamefont {Michel},
  \citenamefont {Noack}, \citenamefont {Paw{\l}owski}, \citenamefont {Pollet},
  \citenamefont {Pruschke}, \citenamefont {Schollw\"ock}, \citenamefont {Todo},
  \citenamefont {Trebst}, \citenamefont {Troyer}, \citenamefont {Werner},\ and\
  \citenamefont {Wessel}}]{alps}%
  \BibitemOpen
  \bibfield  {author} {\bibinfo {author} {\bibfnamefont {A.}~\bibnamefont
  {Albuquerque}}, \bibinfo {author} {\bibfnamefont {F.}~\bibnamefont {Alet}},
  \bibinfo {author} {\bibfnamefont {P.}~\bibnamefont {Corboz}}, \bibinfo
  {author} {\bibfnamefont {P.}~\bibnamefont {Dayal}}, \bibinfo {author}
  {\bibfnamefont {A.}~\bibnamefont {Feiguin}}, \bibinfo {author} {\bibfnamefont
  {S.}~\bibnamefont {Fuchs}}, \bibinfo {author} {\bibfnamefont
  {L.}~\bibnamefont {Gamper}}, \bibinfo {author} {\bibfnamefont
  {E.}~\bibnamefont {Gull}}, \bibinfo {author} {\bibfnamefont {S.}~\bibnamefont
  {G\"urtler}}, \bibinfo {author} {\bibfnamefont {A.}~\bibnamefont {Honecker}},
  \bibinfo {author} {\bibfnamefont {R.}~\bibnamefont {Igarashi}}, \bibinfo
  {author} {\bibfnamefont {M.}~\bibnamefont {K\"orner}}, \bibinfo {author}
  {\bibfnamefont {A.}~\bibnamefont {Kozhevnikov}}, \bibinfo {author}
  {\bibfnamefont {A.}~\bibnamefont {L\"auchli}}, \bibinfo {author}
  {\bibfnamefont {S.}~\bibnamefont {Manmana}}, \bibinfo {author} {\bibfnamefont
  {M.}~\bibnamefont {Matsumoto}}, \bibinfo {author} {\bibfnamefont
  {I.}~\bibnamefont {McCulloch}}, \bibinfo {author} {\bibfnamefont
  {F.}~\bibnamefont {Michel}}, \bibinfo {author} {\bibfnamefont
  {R.}~\bibnamefont {Noack}}, \bibinfo {author} {\bibfnamefont
  {G.}~\bibnamefont {Paw{\l}owski}}, \bibinfo {author} {\bibfnamefont
  {L.}~\bibnamefont {Pollet}}, \bibinfo {author} {\bibfnamefont
  {T.}~\bibnamefont {Pruschke}}, \bibinfo {author} {\bibfnamefont
  {U.}~\bibnamefont {Schollw\"ock}}, \bibinfo {author} {\bibfnamefont
  {S.}~\bibnamefont {Todo}}, \bibinfo {author} {\bibfnamefont {S.}~\bibnamefont
  {Trebst}}, \bibinfo {author} {\bibfnamefont {M.}~\bibnamefont {Troyer}},
  \bibinfo {author} {\bibfnamefont {P.}~\bibnamefont {Werner}}, \ and\ \bibinfo
  {author} {\bibfnamefont {S.}~\bibnamefont {Wessel}},\ }\href@noop {}
  {\bibfield  {journal} {\bibinfo  {journal} {J. Magn. Magn. Mater.}\ }\textbf
  {\bibinfo {volume} {310}},\ \bibinfo {pages} {1187} (\bibinfo {year}
  {2007})}\BibitemShut {NoStop}%
\bibitem [{\citenamefont {Sandvik}(1999)}]{sandvik1999}%
  \BibitemOpen
  \bibfield  {author} {\bibinfo {author} {\bibfnamefont {A.~W.}\ \bibnamefont
  {Sandvik}},\ }\href@noop {} {\bibfield  {journal} {\bibinfo  {journal} {Phys.
  Rev. Lett.}\ }\textbf {\bibinfo {volume} {83}},\ \bibinfo {pages} {3069}
  (\bibinfo {year} {1999})}\BibitemShut {NoStop}%
\bibitem [{\citenamefont {Mazurenko}\ \emph {et~al.}(2006)\citenamefont
  {Mazurenko}, \citenamefont {Mila},\ and\ \citenamefont
  {Anisimov}}]{mazurenko2006}%
  \BibitemOpen
  \bibfield  {author} {\bibinfo {author} {\bibfnamefont {V.~V.}\ \bibnamefont
  {Mazurenko}}, \bibinfo {author} {\bibfnamefont {F.}~\bibnamefont {Mila}}, \
  and\ \bibinfo {author} {\bibfnamefont {V.~I.}\ \bibnamefont {Anisimov}},\
  }\href@noop {} {\bibfield  {journal} {\bibinfo  {journal} {Phys. Rev. B}\
  }\textbf {\bibinfo {volume} {73}},\ \bibinfo {pages} {014418} (\bibinfo
  {year} {2006})}\BibitemShut {NoStop}%
\bibitem [{\citenamefont {Eschrig}\ and\ \citenamefont
  {Koepernik}(2009)}]{wannier}%
  \BibitemOpen
  \bibfield  {author} {\bibinfo {author} {\bibfnamefont {H.}~\bibnamefont
  {Eschrig}}\ and\ \bibinfo {author} {\bibfnamefont {K.}~\bibnamefont
  {Koepernik}},\ }\href@noop {} {\bibfield  {journal} {\bibinfo  {journal}
  {Phys. Rev. B}\ }\textbf {\bibinfo {volume} {80}},\ \bibinfo {pages} {104503}
  (\bibinfo {year} {2009})}\BibitemShut {NoStop}%
\bibitem [{\citenamefont {Garrett}\ \emph {et~al.}(1997)\citenamefont
  {Garrett}, \citenamefont {Nagler}, \citenamefont {Tennant}, \citenamefont
  {Sales},\ and\ \citenamefont {Barnes}}]{garrett1997}%
  \BibitemOpen
  \bibfield  {author} {\bibinfo {author} {\bibfnamefont {A.~W.}\ \bibnamefont
  {Garrett}}, \bibinfo {author} {\bibfnamefont {S.~E.}\ \bibnamefont {Nagler}},
  \bibinfo {author} {\bibfnamefont {D.~A.}\ \bibnamefont {Tennant}}, \bibinfo
  {author} {\bibfnamefont {B.~C.}\ \bibnamefont {Sales}}, \ and\ \bibinfo
  {author} {\bibfnamefont {T.}~\bibnamefont {Barnes}},\ }\href@noop {}
  {\bibfield  {journal} {\bibinfo  {journal} {Phys. Rev. Lett.}\ }\textbf
  {\bibinfo {volume} {79}},\ \bibinfo {pages} {745} (\bibinfo {year}
  {1997})}\BibitemShut {NoStop}%
\bibitem [{\citenamefont {Tennant}\ \emph {et~al.}(1997)\citenamefont
  {Tennant}, \citenamefont {Nagler}, \citenamefont {Garrett}, \citenamefont
  {Barnes},\ and\ \citenamefont {Torardi}}]{tennant1997}%
  \BibitemOpen
  \bibfield  {author} {\bibinfo {author} {\bibfnamefont {D.~A.}\ \bibnamefont
  {Tennant}}, \bibinfo {author} {\bibfnamefont {S.~E.}\ \bibnamefont {Nagler}},
  \bibinfo {author} {\bibfnamefont {A.~W.}\ \bibnamefont {Garrett}}, \bibinfo
  {author} {\bibfnamefont {T.}~\bibnamefont {Barnes}}, \ and\ \bibinfo {author}
  {\bibfnamefont {C.~C.}\ \bibnamefont {Torardi}},\ }\href@noop {} {\bibfield
  {journal} {\bibinfo  {journal} {Phys. Rev. Lett.}\ }\textbf {\bibinfo
  {volume} {78}},\ \bibinfo {pages} {4998} (\bibinfo {year}
  {1997})}\BibitemShut {NoStop}%
\bibitem [{\citenamefont {Nath}\ \emph {et~al.}(2008)\citenamefont {Nath},
  \citenamefont {Kasinathan}, \citenamefont {Rosner}, \citenamefont {Baenitz},\
  and\ \citenamefont {Geibel}}]{nath2008}%
  \BibitemOpen
  \bibfield  {author} {\bibinfo {author} {\bibfnamefont {R.}~\bibnamefont
  {Nath}}, \bibinfo {author} {\bibfnamefont {D.}~\bibnamefont {Kasinathan}},
  \bibinfo {author} {\bibfnamefont {H.}~\bibnamefont {Rosner}}, \bibinfo
  {author} {\bibfnamefont {M.}~\bibnamefont {Baenitz}}, \ and\ \bibinfo
  {author} {\bibfnamefont {C.}~\bibnamefont {Geibel}},\ }\href@noop {}
  {\bibfield  {journal} {\bibinfo  {journal} {Phys. Rev. B}\ }\textbf {\bibinfo
  {volume} {77}},\ \bibinfo {pages} {134451} (\bibinfo {year}
  {2008})}\BibitemShut {NoStop}%
\bibitem [{\citenamefont {Janson}\ \emph {et~al.}(2011)\citenamefont {Janson},
  \citenamefont {Tsirlin}, \citenamefont {Sichelschmidt}, \citenamefont
  {Skourski}, \citenamefont {Weickert},\ and\ \citenamefont
  {Rosner}}]{janson2011}%
  \BibitemOpen
  \bibfield  {author} {\bibinfo {author} {\bibfnamefont {O.}~\bibnamefont
  {Janson}}, \bibinfo {author} {\bibfnamefont {A.~A.}\ \bibnamefont {Tsirlin}},
  \bibinfo {author} {\bibfnamefont {J.}~\bibnamefont {Sichelschmidt}}, \bibinfo
  {author} {\bibfnamefont {Y.}~\bibnamefont {Skourski}}, \bibinfo {author}
  {\bibfnamefont {F.}~\bibnamefont {Weickert}}, \ and\ \bibinfo {author}
  {\bibfnamefont {H.}~\bibnamefont {Rosner}},\ }\href@noop {} {\bibfield
  {journal} {\bibinfo  {journal} {Phys. Rev. B}\ }\textbf {\bibinfo {volume}
  {83}},\ \bibinfo {pages} {094435} (\bibinfo {year} {2011})}\BibitemShut
  {NoStop}%
\bibitem [{\citenamefont {R\"uegg}\ \emph {et~al.}(2007)\citenamefont
  {R\"uegg}, \citenamefont {McMorrow}, \citenamefont {Normand}, \citenamefont
  {R{\o}nnow}, \citenamefont {Sebastian}, \citenamefont {Fisher}, \citenamefont
  {Batista}, \citenamefont {Gvasaliya}, \citenamefont {Niedermayer},\ and\
  \citenamefont {Stahn}}]{ruegg2007}%
  \BibitemOpen
  \bibfield  {author} {\bibinfo {author} {\bibfnamefont {C.}~\bibnamefont
  {R\"uegg}}, \bibinfo {author} {\bibfnamefont {D.~F.}\ \bibnamefont
  {McMorrow}}, \bibinfo {author} {\bibfnamefont {B.}~\bibnamefont {Normand}},
  \bibinfo {author} {\bibfnamefont {H.~M.}\ \bibnamefont {R{\o}nnow}}, \bibinfo
  {author} {\bibfnamefont {S.~E.}\ \bibnamefont {Sebastian}}, \bibinfo {author}
  {\bibfnamefont {I.~R.}\ \bibnamefont {Fisher}}, \bibinfo {author}
  {\bibfnamefont {C.~D.}\ \bibnamefont {Batista}}, \bibinfo {author}
  {\bibfnamefont {S.~N.}\ \bibnamefont {Gvasaliya}}, \bibinfo {author}
  {\bibfnamefont {C.}~\bibnamefont {Niedermayer}}, \ and\ \bibinfo {author}
  {\bibfnamefont {J.}~\bibnamefont {Stahn}},\ }\href@noop {} {\bibfield
  {journal} {\bibinfo  {journal} {Phys. Rev. Lett.}\ }\textbf {\bibinfo
  {volume} {98}},\ \bibinfo {pages} {017202} (\bibinfo {year}
  {2007})}\BibitemShut {NoStop}%
\bibitem [{\citenamefont {Mazurenko}\ \emph {et~al.}()\citenamefont
  {Mazurenko}, \citenamefont {Valentyuk}, \citenamefont {Stern},\ and\
  \citenamefont {Tsirlin}}]{bacusi2o6}%
  \BibitemOpen
  \bibfield  {author} {\bibinfo {author} {\bibfnamefont {V.~V.}\ \bibnamefont
  {Mazurenko}}, \bibinfo {author} {\bibfnamefont {M.~V.}\ \bibnamefont
  {Valentyuk}}, \bibinfo {author} {\bibfnamefont {R.}~\bibnamefont {Stern}}, \
  and\ \bibinfo {author} {\bibfnamefont {A.~A.}\ \bibnamefont {Tsirlin}},\
  }\href@noop {} {}\bibinfo {note} {{a}rXiv:1309.6762}\BibitemShut {NoStop}%
\bibitem [{\citenamefont {Johnston}\ \emph {et~al.}(2000)\citenamefont
  {Johnston}, \citenamefont {Kremer}, \citenamefont {Troyer}, \citenamefont
  {Wang}, \citenamefont {Kl\"umper}, \citenamefont {Bud'ko}, \citenamefont
  {Panchula},\ and\ \citenamefont {Canfield}}]{johnston2000}%
  \BibitemOpen
  \bibfield  {author} {\bibinfo {author} {\bibfnamefont {D.~C.}\ \bibnamefont
  {Johnston}}, \bibinfo {author} {\bibfnamefont {R.~K.}\ \bibnamefont
  {Kremer}}, \bibinfo {author} {\bibfnamefont {M.}~\bibnamefont {Troyer}},
  \bibinfo {author} {\bibfnamefont {X.}~\bibnamefont {Wang}}, \bibinfo {author}
  {\bibfnamefont {A.}~\bibnamefont {Kl\"umper}}, \bibinfo {author}
  {\bibfnamefont {S.~L.}\ \bibnamefont {Bud'ko}}, \bibinfo {author}
  {\bibfnamefont {A.~F.}\ \bibnamefont {Panchula}}, \ and\ \bibinfo {author}
  {\bibfnamefont {P.~C.}\ \bibnamefont {Canfield}},\ }\href@noop {} {\bibfield
  {journal} {\bibinfo  {journal} {Phys. Rev. B}\ }\textbf {\bibinfo {volume}
  {61}},\ \bibinfo {pages} {9558} (\bibinfo {year} {2000})}\BibitemShut
  {NoStop}%
\bibitem [{\citenamefont {Ivanshin}\ \emph {et~al.}(2003)\citenamefont
  {Ivanshin}, \citenamefont {Yushankhai}, \citenamefont {Sichelschmidt},
  \citenamefont {Zakharov}, \citenamefont {Kaul},\ and\ \citenamefont
  {Geibel}}]{ivanshin2003}%
  \BibitemOpen
  \bibfield  {author} {\bibinfo {author} {\bibfnamefont {V.~A.}\ \bibnamefont
  {Ivanshin}}, \bibinfo {author} {\bibfnamefont {V.}~\bibnamefont
  {Yushankhai}}, \bibinfo {author} {\bibfnamefont {J.}~\bibnamefont
  {Sichelschmidt}}, \bibinfo {author} {\bibfnamefont {D.~V.}\ \bibnamefont
  {Zakharov}}, \bibinfo {author} {\bibfnamefont {E.~E.}\ \bibnamefont {Kaul}},
  \ and\ \bibinfo {author} {\bibfnamefont {C.}~\bibnamefont {Geibel}},\
  }\href@noop {} {\bibfield  {journal} {\bibinfo  {journal} {Phys. Rev. B}\
  }\textbf {\bibinfo {volume} {68}},\ \bibinfo {pages} {064404} (\bibinfo
  {year} {2003})}\BibitemShut {NoStop}%
\bibitem [{\citenamefont {F{\"o}rster}\ \emph {et~al.}(2013)\citenamefont
  {F{\"o}rster}, \citenamefont {Garcia}, \citenamefont {Gruner}, \citenamefont
  {Kaul}, \citenamefont {Schmidt}, \citenamefont {Geibel},\ and\ \citenamefont
  {Sichelschmidt}}]{foerster2013}%
  \BibitemOpen
  \bibfield  {author} {\bibinfo {author} {\bibfnamefont {T.}~\bibnamefont
  {F{\"o}rster}}, \bibinfo {author} {\bibfnamefont {F.~A.}\ \bibnamefont
  {Garcia}}, \bibinfo {author} {\bibfnamefont {T.}~\bibnamefont {Gruner}},
  \bibinfo {author} {\bibfnamefont {E.~E.}\ \bibnamefont {Kaul}}, \bibinfo
  {author} {\bibfnamefont {B.}~\bibnamefont {Schmidt}}, \bibinfo {author}
  {\bibfnamefont {C.}~\bibnamefont {Geibel}}, \ and\ \bibinfo {author}
  {\bibfnamefont {J.}~\bibnamefont {Sichelschmidt}},\ }\href@noop {} {\bibfield
   {journal} {\bibinfo  {journal} {Phys. Rev. B}\ }\textbf {\bibinfo {volume}
  {87}},\ \bibinfo {pages} {180401(R)} (\bibinfo {year} {2013})}\BibitemShut
  {NoStop}%
\bibitem [{Note2()}]{Note2}%
  \BibitemOpen
  \bibinfo {note} {Each [VOV$_2$O$_7$] layer is shifted for $b/2$ with respect
  to its neighboring layers. Therefore, each V$^{+4}$ ion interacts with two
  V$^{+4}$ ions from the neighboring layer, which then leads to a triangle of
  frustrated AFM interactions $J_{\perp \alpha }-J_b-J_{\perp \alpha }$
  ($\alpha =1,2$). While on the classical level this frustration would prevent
  the system from any 3D magnetic order, quantum fluctuations and/or magnetic
  anisotropy will typically induce the long-range order at low temperatures.
  Considering the low energies of interlayer couplings and relevant
  anisotropies, this coupling regime is quite challenging for a quantitative
  microscopic modeling.}\BibitemShut {Stop}%
\bibitem [{\citenamefont {Binder}(1997)}]{binder1997}%
  \BibitemOpen
  \bibfield  {author} {\bibinfo {author} {\bibfnamefont {K.}~\bibnamefont
  {Binder}},\ }\href@noop {} {\bibfield  {journal} {\bibinfo  {journal} {Rep.
  Prog. Phys.}\ }\textbf {\bibinfo {volume} {60}},\ \bibinfo {pages} {487}
  (\bibinfo {year} {1997})}\BibitemShut {NoStop}%
\bibitem [{\citenamefont {Affleck}\ and\ \citenamefont
  {Oshikawa}(1999)}]{affleck1999}%
  \BibitemOpen
  \bibfield  {author} {\bibinfo {author} {\bibfnamefont {I.}~\bibnamefont
  {Affleck}}\ and\ \bibinfo {author} {\bibfnamefont {M.}~\bibnamefont
  {Oshikawa}},\ }\href@noop {} {\bibfield  {journal} {\bibinfo  {journal}
  {Phys. Rev. B}\ }\textbf {\bibinfo {volume} {60}},\ \bibinfo {pages} {1038}
  (\bibinfo {year} {1999})}\BibitemShut {NoStop}%
\bibitem [{\citenamefont {Asano}\ \emph {et~al.}(2000)\citenamefont {Asano},
  \citenamefont {Nojiri}, \citenamefont {Inagaki}, \citenamefont {Boucher},
  \citenamefont {Sakon}, \citenamefont {Ajiro},\ and\ \citenamefont
  {Motokawa}}]{asano2000}%
  \BibitemOpen
  \bibfield  {author} {\bibinfo {author} {\bibfnamefont {T.}~\bibnamefont
  {Asano}}, \bibinfo {author} {\bibfnamefont {H.}~\bibnamefont {Nojiri}},
  \bibinfo {author} {\bibfnamefont {Y.}~\bibnamefont {Inagaki}}, \bibinfo
  {author} {\bibfnamefont {J.~P.}\ \bibnamefont {Boucher}}, \bibinfo {author}
  {\bibfnamefont {T.}~\bibnamefont {Sakon}}, \bibinfo {author} {\bibfnamefont
  {Y.}~\bibnamefont {Ajiro}}, \ and\ \bibinfo {author} {\bibfnamefont
  {M.}~\bibnamefont {Motokawa}},\ }\href@noop {} {\bibfield  {journal}
  {\bibinfo  {journal} {Phys. Rev. Lett.}\ }\textbf {\bibinfo {volume} {84}},\
  \bibinfo {pages} {5880} (\bibinfo {year} {2000})}\BibitemShut {NoStop}%
\bibitem [{\citenamefont {Moriya}(1960)}]{moriya1960}%
  \BibitemOpen
  \bibfield  {author} {\bibinfo {author} {\bibfnamefont {T.}~\bibnamefont
  {Moriya}},\ }\href@noop {} {\bibfield  {journal} {\bibinfo  {journal} {Phys.
  Rev.}\ }\textbf {\bibinfo {volume} {120}},\ \bibinfo {pages} {91} (\bibinfo
  {year} {1960})}\BibitemShut {NoStop}%
\bibitem [{Note3()}]{Note3}%
  \BibitemOpen
  \bibinfo {note} {Magnetic anisotropy parameters are extracted from total
  energies, which are converged to $10^{-6}$\protect \,eV/f.u. However,
  numerical issues, including the $k$-point integration, lead to a somewhat
  lower accuracy that we estimate as $10^{-5}$\protect \,eV/f.u. (i.e., about
  0.05\protect \,K for individual magnetic couplings) according to the mismatch
  between total energies of same spin configurations calculated in different
  supercells.}\BibitemShut {Stop}%
\bibitem [{\citenamefont {Tsirlin}\ \emph {et~al.}(2010)\citenamefont
  {Tsirlin}, \citenamefont {Janson},\ and\ \citenamefont
  {Rosner}}]{tsirlin2010}%
  \BibitemOpen
  \bibfield  {author} {\bibinfo {author} {\bibfnamefont {A.~A.}\ \bibnamefont
  {Tsirlin}}, \bibinfo {author} {\bibfnamefont {O.}~\bibnamefont {Janson}}, \
  and\ \bibinfo {author} {\bibfnamefont {H.}~\bibnamefont {Rosner}},\
  }\href@noop {} {\bibfield  {journal} {\bibinfo  {journal} {Phys. Rev. B}\
  }\textbf {\bibinfo {volume} {82}},\ \bibinfo {pages} {144416} (\bibinfo
  {year} {2010})}\BibitemShut {NoStop}%
\bibitem [{\citenamefont {Sa\'ul}\ and\ \citenamefont
  {Radtke}(2011)}]{saul2011}%
  \BibitemOpen
  \bibfield  {author} {\bibinfo {author} {\bibfnamefont {A.}~\bibnamefont
  {Sa\'ul}}\ and\ \bibinfo {author} {\bibfnamefont {G.}~\bibnamefont
  {Radtke}},\ }\href@noop {} {\bibfield  {journal} {\bibinfo  {journal} {Phys.
  Rev. Lett.}\ }\textbf {\bibinfo {volume} {106}},\ \bibinfo {pages} {177203}
  (\bibinfo {year} {2011})}\BibitemShut {NoStop}%
\bibitem [{\citenamefont {Tsirlin}\ \emph {et~al.}(2012)\citenamefont
  {Tsirlin}, \citenamefont {M\"oller}, \citenamefont {Lorenz}, \citenamefont
  {Skourski},\ and\ \citenamefont {Rosner}}]{tsirlin2012}%
  \BibitemOpen
  \bibfield  {author} {\bibinfo {author} {\bibfnamefont {A.~A.}\ \bibnamefont
  {Tsirlin}}, \bibinfo {author} {\bibfnamefont {A.}~\bibnamefont {M\"oller}},
  \bibinfo {author} {\bibfnamefont {B.}~\bibnamefont {Lorenz}}, \bibinfo
  {author} {\bibfnamefont {Y.}~\bibnamefont {Skourski}}, \ and\ \bibinfo
  {author} {\bibfnamefont {H.}~\bibnamefont {Rosner}},\ }\href@noop {}
  {\bibfield  {journal} {\bibinfo  {journal} {Phys. Rev. B}\ }\textbf {\bibinfo
  {volume} {85}},\ \bibinfo {pages} {014401} (\bibinfo {year}
  {2012})}\BibitemShut {NoStop}%
\bibitem [{\citenamefont {Belik}\ \emph {et~al.}(2005)\citenamefont {Belik},
  \citenamefont {Azuma}, \citenamefont {Matsuo}, \citenamefont {Kindo},\ and\
  \citenamefont {Takano}}]{belik2005}%
  \BibitemOpen
  \bibfield  {author} {\bibinfo {author} {\bibfnamefont {A.~A.}\ \bibnamefont
  {Belik}}, \bibinfo {author} {\bibfnamefont {M.}~\bibnamefont {Azuma}},
  \bibinfo {author} {\bibfnamefont {A.}~\bibnamefont {Matsuo}}, \bibinfo
  {author} {\bibfnamefont {K.}~\bibnamefont {Kindo}}, \ and\ \bibinfo {author}
  {\bibfnamefont {M.}~\bibnamefont {Takano}},\ }\href@noop {} {\bibfield
  {journal} {\bibinfo  {journal} {Inorg. Chem.}\ }\textbf {\bibinfo {volume}
  {44}},\ \bibinfo {pages} {3762} (\bibinfo {year} {2005})}\BibitemShut
  {NoStop}%
\bibitem [{\citenamefont {Shpanchenko}\ \emph {et~al.}(2003)\citenamefont
  {Shpanchenko}, \citenamefont {Chernaya}, \citenamefont {Antipov},
  \citenamefont {Hadermann}, \citenamefont {Kaul},\ and\ \citenamefont
  {Geibel}}]{shpanchenko2003}%
  \BibitemOpen
  \bibfield  {author} {\bibinfo {author} {\bibfnamefont {R.~V.}\ \bibnamefont
  {Shpanchenko}}, \bibinfo {author} {\bibfnamefont {V.~V.}\ \bibnamefont
  {Chernaya}}, \bibinfo {author} {\bibfnamefont {E.~V.}\ \bibnamefont
  {Antipov}}, \bibinfo {author} {\bibfnamefont {J.}~\bibnamefont {Hadermann}},
  \bibinfo {author} {\bibfnamefont {E.~E.}\ \bibnamefont {Kaul}}, \ and\
  \bibinfo {author} {\bibfnamefont {C.}~\bibnamefont {Geibel}},\ }\href@noop {}
  {\bibfield  {journal} {\bibinfo  {journal} {J. Solid State Chem.}\ }\textbf
  {\bibinfo {volume} {173}},\ \bibinfo {pages} {244} (\bibinfo {year}
  {2003})}\BibitemShut {NoStop}%
\bibitem [{Note4()}]{Note4}%
  \BibitemOpen
  \bibinfo {note} {We use same parameters as for BaV$_3$O$_8$: GGA+$U$,
  $U_d=4$\protect \,eV, $J_d=1$\protect \,eV, fully-localized-limit
  double-counting correction. Total energies are obtained on a $k$ mesh with 64
  points in the first Brillouin zone.}\BibitemShut {Stop}%
\bibitem [{\citenamefont {Cao}\ \emph {et~al.}(2005)\citenamefont {Cao},
  \citenamefont {Haraldsen}, \citenamefont {Brown}, \citenamefont {Musfeldt},
  \citenamefont {Thompson}, \citenamefont {Zvyagin}, \citenamefont {Krzystek},
  \citenamefont {Whangbo}, \citenamefont {Nagler},\ and\ \citenamefont
  {Torardi}}]{cao2005}%
  \BibitemOpen
  \bibfield  {author} {\bibinfo {author} {\bibfnamefont {J.}~\bibnamefont
  {Cao}}, \bibinfo {author} {\bibfnamefont {J.~T.}\ \bibnamefont {Haraldsen}},
  \bibinfo {author} {\bibfnamefont {S.}~\bibnamefont {Brown}}, \bibinfo
  {author} {\bibfnamefont {J.~L.}\ \bibnamefont {Musfeldt}}, \bibinfo {author}
  {\bibfnamefont {J.~R.}\ \bibnamefont {Thompson}}, \bibinfo {author}
  {\bibfnamefont {S.}~\bibnamefont {Zvyagin}}, \bibinfo {author} {\bibfnamefont
  {J.}~\bibnamefont {Krzystek}}, \bibinfo {author} {\bibfnamefont {M.-H.}\
  \bibnamefont {Whangbo}}, \bibinfo {author} {\bibfnamefont {S.~E.}\
  \bibnamefont {Nagler}}, \ and\ \bibinfo {author} {\bibfnamefont {C.~C.}\
  \bibnamefont {Torardi}},\ }\href@noop {} {\bibfield  {journal} {\bibinfo
  {journal} {Phys. Rev. B}\ }\textbf {\bibinfo {volume} {72}},\ \bibinfo
  {pages} {214421} (\bibinfo {year} {2005})}\BibitemShut {NoStop}%
\end{thebibliography}
%

\clearpage
\renewcommand{\thefigure}{S\arabic{figure}}
\setcounter{figure}{0}
\setcounter{table}{1}
\renewcommand{\thetable}{S\arabic{table}}
\renewcommand{\theequation}{S\arabic{equation}}

\begin{widetext}
	\begin{center}
{\large
Supplemental material for
\smallskip

\textbf{Spin-chain magnetism and uniform Dzyloshinsky-Moriya anisotropy in BaV$_3$O$_8$}
}

\medskip

Alexander A. Tsirlin
\end{center}
\end{widetext}
\medskip

\subsubsection{Evaluation of magnetic couplings using the four-configuration method}

Exchange couplings in insulators are typically obtained from total energies of different spin configurations. This approach is based on the classical treatment of the spin Hamiltonian (Eq.~(1) of the manuscript) that provides the energy of an arbitrary spin configuration as a linear combination of magnetic couplings $J_i$, $\Dv_i$, etc. Let's first restrict ourselves to isotropic couplings $J_i$. A system with $N$ magnetic couplings ($i=1,\ldots N$) requires the evaluation of total energies for $N+1$ spin configurations, because our model expressions for the energy contain $N$ parameters $J_i$ and the reference energy $E_0$, which is the energy of the non-magnetic state. This method is known as the supercell approach or mapping analysis.

\begin{figure}
\centerline{\includegraphics{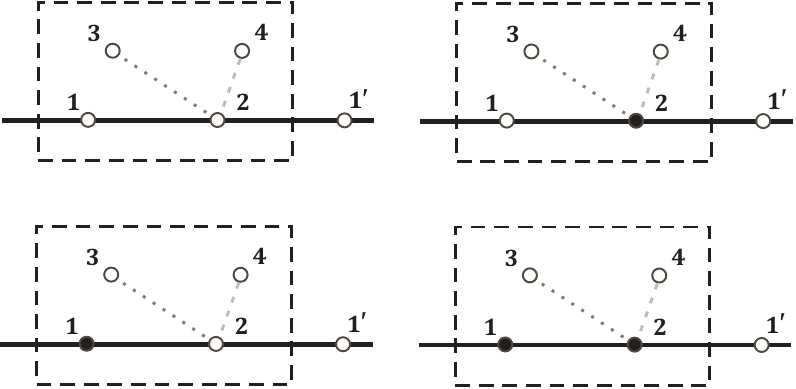}}
\caption{\label{fig:configurations}
Different spin configurations used in the mapping analysis and in the four-configuration method. The two upper configurations, $\uparrow\uparrow\uparrow\uparrow$ and $\uparrow\downarrow\uparrow\uparrow$, yield a linear combination of several couplings, $J_{12}+J_{23}+J_{24}+J_{1'2}$. Four configurations remove the linear dependence between $J_{12}$ and $J_{23},J_{24}$, but the linear dependence between $J_{12}$ and $J_{1'2}$ is caused by the translation symmetry and can not be removed without expanding the unit cell along the $1-2-1'$ direction.
}
\end{figure}

Alternatively, we can focus on a single coupling $J_{pq}$ between spins $p$ and $q$, and use four spin configurations, where spins $p$ and $q$ have all possible arrangements ($\uparrow\uparrow$, $\uparrow\downarrow$, $\downarrow\uparrow$, $\downarrow\downarrow$), whereas other spins are fixed in any (arbitrary) configuration (see Fig.~\ref{fig:configurations}). According to Xiang \textit{et al.},\cite{xiang2011} a linear combination of these four energies yields $J_{pq}$, and all other couplings are cancelled:
\begin{equation}
  J_{pq}=\frac{1}{4S^2}(E_{\uparrow\uparrow}-E_{\uparrow\downarrow}-E_{\downarrow\uparrow}+ E_{\downarrow\downarrow}).
\label{eq:four-spin}\end{equation}   
This approach requires that the energies of $4N$ spin configurations are calculated to obtain $N$ couplings $J_i$. Computationally, this approach is rather inefficient, but it largely simplifies the analysis of complex magnets with multiple, heavily intertwined exchanges $J_i$.

Let's compare these two methods for a simple system with 4 magnetic atoms in the unit cell. Suppose we want to obtain the coupling $J_{12}$. The standard mapping analysis can be done with only two spin configurations, and naively 
\begin{equation}
  2J_{12}S^2=E_{\uparrow\uparrow\uparrow\uparrow}-E_{\uparrow\downarrow\uparrow\uparrow},
\end{equation}  
but in fact other couplings may be present as well. The complete expression for the energy difference is:
\begin{equation}
  E_{\uparrow\uparrow\uparrow\uparrow}-E_{\uparrow\downarrow\uparrow\uparrow}=2(J_{12}+J_{23}+J_{24}+J_{1'2})S^2,
\end{equation}    
where we took into account that atoms $1$ and $1'$ are related by symmetry, hence the coupling $J_{1'2}$ is \textit{always} added to $J_{12}$.

The four-configuration approach solves one of the problems. It disentangles $J_{12}$ from other couplings within the same unit cell:
\begin{eqnarray*}
  E_{\uparrow\uparrow\uparrow\uparrow}=E_0+(J_{12}+J_{1'2}+J_{23}+J_{24})S^2, \\
  E_{\uparrow\downarrow\uparrow\uparrow}=E_0+(-J_{12}-J_{1'2}-J_{23}-J_{24})S^2, \\
  E_{\downarrow\uparrow\uparrow\uparrow}=E_0+(-J_{12}-J_{1'2}+J_{23}+J_{24})S^2, \\
  E_{\downarrow\downarrow\uparrow\uparrow}=E_0+(J_{12}+J_{1'2}-J_{23}-J_{24})S^2. 
\end{eqnarray*}
Then,
\begin{equation}
  4(J_{12}+J_{1'2})S^2=E_{\uparrow\uparrow\uparrow\uparrow}-E_{\uparrow\downarrow\uparrow\uparrow}- E_{\downarrow\uparrow\uparrow\uparrow}+E_{\downarrow\downarrow\uparrow\uparrow}.
\end{equation}  
However, this four-configuration approach does not break the relation between $J_{12}$ and $J_{1'2}$, because the atoms $1$ and $1'$ remain translationally equivalent.

The four-configuration approach is particularly useful in situations when the presence of multiple exchange couplings can be envisaged. The ambiguity related to the translation symmetry is removed by increasing the unit cell. For example, in a doubled supercell atoms $1$ and $1'$ are no longer equivalent, hence $J_{12}$ and $J_{1'2}$ can be obtained independently.

The evaluation of $J_i$ from total energies can be easily generalized to magnetic anisotropy parameters by including the spin-orbit coupling and choosing the direction of spins. For example, Eq.~(\ref{eq:four-spin}) with spins directed along $x$, $y$, and $z$ yields diagonal terms of the symmetric anisotropy tensor, $J_{pq}^{xx}$, $J_{pq}^{yy}$, and $J_{pq}^{zz}$, respectivly. The DM interactions couple different component of spins $p$ and $q$. Therefore, we should direct spin $p$ along, e.g., $x$ and spin $q$ along $y$ to obtain the $z$-component of the DM vector $\Dv_{pq}$.

The four-configuration approach becomes especially useful for magnetic anisotropy parameters that change their signs and directions when going from one bond to another. It is usually more convenient to focus on a single bond than to analyze all symmetry relations in a complex crystal structure. The problem of the translation symmetry should not be ignored, though. In a smaller unit cell with $1$ equivalent to $1'$, the coupling $12$ is always augmented by the coupling $1'2$. For the DM vectors, the calculations in a small unit cell will yield $\Dv_{12}+\Dv_{1'2}=0$ when the DM anisotropy is uniform, as in BaV$_3$O$_8$, where $\Dv_{12}=\Dv_{21'}=-\Dv_{1'2}$. This problem necessitates us to use 4-fold supercells (96 atoms) for the evaluation of the DM vectors. Doubled unit cells with 48 atoms are not sufficient to evaluate the DM vectors in BaV$_3$O$_8$.

\subsubsection{Symmetry relations. Application to VOHPO$_4\cdot 0.5$H$_2$O}
\begin{figure}
\centerline{\includegraphics{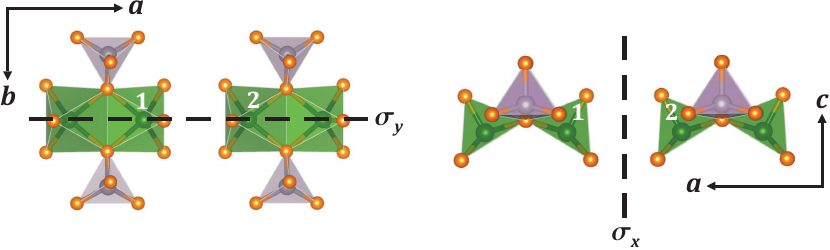}}
\caption{\label{fig:vpo}
Crystal structure of VOHPO$_4\cdot 0.5$H$_2$O showing the two mirror planes: $\sigma_x$ runs between the interacting V$^{+4}$ ions 1 and 2, thus removing the $D_x$ component of the DM vector; $\sigma_y$ contains both 1 and 2 and requires that $\Dv$ is perpendicular to the $ac$ plane, i.e., $\Dv\|\mathbf b$.
}
\end{figure}
Symmetry arguments require that $\Dv_{12}=-\Dv_{21}$ and, given a sufficiently high symmetry, certain components of the DM vectors are zero. These constraints can be used to validate our computational approach and make sure that the calculated DM vectors, despite being diminutively small (typically, below 1~meV), are reliable. Here, we do it for VOHPO$_4\cdot 0.5$H$_2$O, a system of weakly coupled magnetic dimers residing between the structural dimers (Fig.~\ref{fig:vpo}).\footnote{We use same parameters as for BaV$_3$O$_8$: GGA+$U$, $U_d=4$\,eV, $J_d=1$\,eV, fully-localized-limit double-counting correction. Total energies are obtained on a $k$ mesh with 64 points in the first Brillouin zone.} 

In VOHPO$_4\cdot 0.5$H$_2$O, the intradimer coupling connects two V$^{4+}$ ions related by the mirror symmetry ($\sigma_x$ plane perpendicular to the $a$ axis). Therefore, the DM vector should lie in this plane ($D_x=0$).  Additionally, both V$^{+4}$ ions lie in the $\sigma_y$ mirror plane, which then implies the DM vector perpendicular to this plane ($D_y\neq 0$, $D_z=0$). Applying GGA+$U$ calculations and the four-configuration method, we find (in units of~K):
\begin{eqnarray*}
\Dv_{12}=(0.008,-0.57,-0.008),\\
\Dv_{21}=(0.002,0.58,0.003).
\end{eqnarray*}
Indeed, the largest component of the $\Dv$ vector is along the $b$ direction. The $D_x$ and $D_z$ components are zero within the error bar. This error bar is related to the accuracy of individual total energies. We converge the energy to $\sim 10^{-6}$\,eV/cell (about 0.01\,K) in each calculation, whereas the actual precision (e.g., the reproducibility between different unit cells) is lower, about $10^{-5}$\,eV/cell or about 0.05\,K for a single magnetic coupling. We then conclude that $\Dv_{12}\simeq (0,0.57,0)$ and $\Dv_{21}=-\Dv_{12}$ in full agreement with the symmetry analysis.

We have also computed the isotropic exchange $J\simeq 71$\,K (compare to the experimental\cite{tennant1997} $J\simeq 91$\,K) and evaluated $|\Dv|/J\simeq 0.008$, which is of the correct order of magnitude, but lower than the available experimental estimate of $|\Dv|/J\simeq 0.02$, see Ref.~\onlinecite{cao2005}. Extensive benchmark tests for a broader range of compounds are presently underway and will be published elsewhere. Nevertheless, already our first results demonstrate the excellent performance of the total-energy, four-configuration approach to the evaluation of DM couplings in insulators.

\end{document}